\documentclass[floatfix,%
 reprint,fleqn,
 amsmath,amssymb,
 aps,
 prl
]{revtex4-2}

\usepackage{amssymb,amsmath,amstext}
\usepackage{graphicx}
\usepackage{bm}
\usepackage{hyperref}
\usepackage{xcolor}

\usepackage{physics}
\usepackage{ulem}
\usepackage{dcolumn}
\usepackage[caption=false]{subfig}
\usepackage{hyperref}
\usepackage[T1]{fontenc}

\begin{document}

\preprint{APS/123-QED}

\title{Confinement Reveals Hidden Splay-Bend Order in Twist-Bend Nematics}%

\author{Szymon Drzazga$^{1}$}%
\affiliation{$^{1}$Institute of Theoretical Physics and  Mark Kac Center for Complex Systems Research, Jagiellonian University, \L{}ojasiewicza 11, 30-348 Krak\'ow, Poland}%
\email{szymon.drzazga@doctoral.uj.edu.pl}%
\author{Piotr Kubala$^{1}$}%
\affiliation{$^{1}$Institute of Theoretical Physics and  Mark Kac Center for Complex Systems Research, Jagiellonian University, \L{}ojasiewicza 11, 30-348 Krak\'ow, Poland}%
\email{piotr.kubala@doctoral.uj.edu.pl}%

\author{Lech Longa$^{1,2}$}%
\affiliation{$^{1}$Institute of Theoretical Physics and  Mark Kac Center for Complex Systems Research, Jagiellonian University, \L{}ojasiewicza 11, 30-348 Krak\'ow, Poland}%
\affiliation{$^2$International Institute for Sustainability with Knotted Chiral Meta Matter (WPI-SKCM²),
Hiroshima University, 1-3-1 Kagamiyama, Higashi-Hiroshima, Hiroshima 739-8526, Japan}%
\email{lech.longa@uj.edu.pl}%

\date{\today}

\begin{abstract}
Using extensive Monte Carlo (MC) and molecular dynamics (MD) simulations, we
investigate how spatial confinement affects molecular organization within thin
films of the nematic twist-bend ($\mathrm{N_{TB}}$) phase. Our simulations show
that confinement markedly amplifies the otherwise elusive splay-bend order,
primarily by suppressing the intrinsic three-dimensional heliconical structure
characteristic of bulk $\mathrm{N_{TB}}$. 
Remarkably, when the $\mathrm{N_{TB}}$ phase is confined between parallel walls
imposing planar anchoring, and the bulk wave vector is oriented parallel to the
walls, a smectic splay-bend ($\mathrm{S_{SB}}$) phase spontaneously emerges near
the confining surfaces. This intermediate structure subsequently transforms into
the bulk $\mathrm{N_{TB}}$ phase either directly via a smectic splay-bend-twist
($\mathrm{S_{SBT}}$) phase or through a sequence involving both the
$\mathrm{S_{SBT}}$ and the nematic splay-bend-twist ($\mathrm{N_{SBT}}$)
phases. Notably, the $\mathrm{N_{SBT}}$ phase becomes particularly pronounced as
the molecular bend angle approaches its maximum attainable value in bulk
$\mathrm{N_{TB}}$; this regime occurs in close proximity to the
$\mathrm{\mathrm{N_{TB}{-}S_{A}}}$ transition line on the
bulk phase diagram.
Our findings reveal a compelling and intricate interplay among chirality,
confinement, and molecular ordering, further evidenced by the calculated
elementary director distortions. Crucially, this study opens promising avenues
for experimental exploration: confined thin-film geometries serve 
as powerful model systems for
revealing and characterizing novel nematic and smectic liquid-crystal phases that remain elusive in, or
currently inaccessible to, bulk experiments.
\end{abstract}
\maketitle

\textit{Introduction.---}
Over the past decade, significant advances in liquid-crystal research have been 
driven by the discovery of novel polar nematic phases \cite{Cestari2011,Borshch2013,Chen2013,Jakli2018,%
PhysRevX.8.041025,PhysRevLett.124.037801,chen2020first,Mandle2021,Mandle2022polar,Fernandez-Rico2020,Kotni2022,Mertelj2018,Chen2020}. 
In these phases, 
molecules exhibit diverse forms of long-range orientational and polar order, 
while their centers of mass remain randomly distributed, as in isotropic fluids.
Among the most striking examples are the twist-bend nematic ($\mathrm{N_{TB}}$) \cite{Cestari2011,Borshch2013,Chen2013,RevModPhys.90.045004,Mandle2022} and the
splay-bend  nematic ($\mathrm{N_{SB}}$) \cite{Fernandez-Rico2020,Kotni2022} phases, 
realized in systems of chemically achiral bent-core molecules and
colloids.
\begin{figure}\label{fig:sketch_of_phases}
    \centering
    \includegraphics[width=1.3\linewidth]{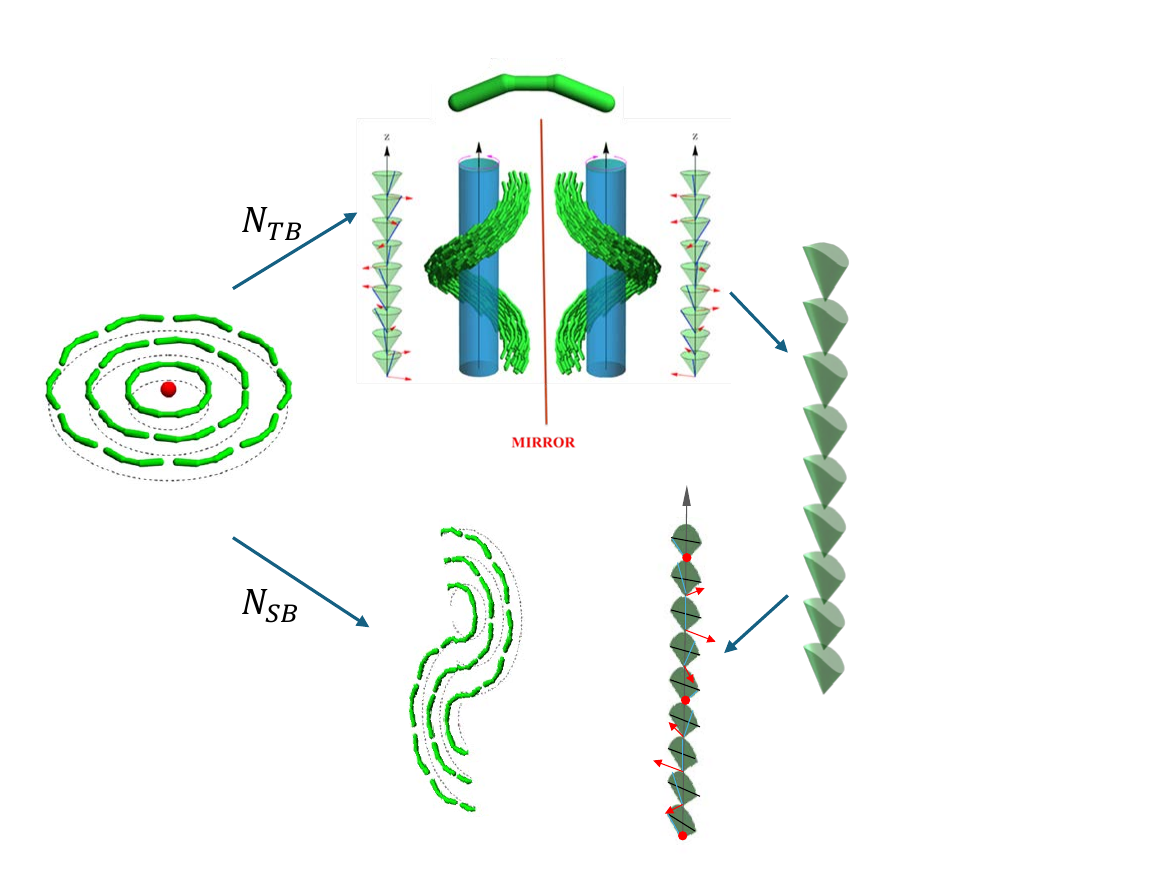}
   \caption{Illustration of how the experimentally observed reduction of the bend elastic constant 
   (\(K_{33}\))—typically the largest Frank elastic constant—drives bent-core mesogens to self-organize into
   periodically modulated nematic structures, most notably the ambidextrous chiral twist-bend (\(\mathrm{N_{TB}}\)) 
   and the biaxial nonchiral splay-bend (\(\mathrm{N_{SB}}\)) phases. 
\textit{\textbf{Left panel:}} Planar bend deformations are favored when \(K_{33}\!\approx\!0\) but,
if extended over macroscopic distances, incur costly defects (red dot).
\textit{\textbf{Center panel:}} ``Escape into the third dimension'' relieves defect-induced
frustration, stabilizing the heliconical \(\mathrm{N_{TB}}\) phase. Alternatively, a nonchiral
\(\mathrm{N_{SB}}\) phase with alternating planar splay--bend regions may form. Red arrows
indicate local polarization; blue rods depict the director. More complex \(\mathrm{N_{SB}}\)-like
textures can also arise from twist-free director fields constructed as normalized gradients
of scalar fields.
\textit{\textbf{Right panel:}} Hybrid splay--bend--twist nematics (\(\mathrm{N_{SBT}}\)) can emerge
by combining splay, bend, and twist deformations, thereby interpolating between \(\mathrm{N_{SB}}\)
and \(\mathrm{N_{TB}}\).}
    \label{fig:NTB_NSB}
\end{figure}

In the \(N_{TB}\) phase, achiral bent-shaped mesogens form a heliconical, locally polar nematic
(Fig.~\ref{fig:NTB_NSB}). The primary order parameter is a transverse vector polarization
\(\mathbf p(z)\), orthogonal to both the director \(\mathbf n(z)\) and
the helical axis (parallel to the  wavevector \(\mathbf{k}\)). The texture combines 
twist–bend distortions of \(\mathbf n\)
with a co-precessing polarization. The local point symmetry is
chiral, polar monoclinic \(C_2\), with the twofold (polar) axis parallel to
\(\mathbf p\). The director maintains a constant tilt and, together with
\(\mathbf p\), precesses with a single pitch (\(  \frac{2 \pi}{k}\)), typically
on the order of \(\sim 10\,\mathrm{nm}\).

Formation of the $\mathrm{N_{TB}}$ phase requires no molecular chirality, yielding
equally probable left- and right-handed heliconical domains. The weakly
first-order transition from the uniaxial nematic ($\mathrm{N_U}$) or isotropic
($\mathrm{Iso}$) phase to $\mathrm{N_{TB}}$ constitutes spontaneous
mirror-symmetry breaking in the absence of long-range translational order. By
contrast, the nonchiral and globally nonpolar $\mathrm{N_{SB}}$ phase exhibits
periodic splay--bend modulations of the director and polarization fields
confined to a single plane, with the polarization usually either perpendicular to the
local director or locally vanishing (Fig.~\ref{fig:NTB_NSB}).

From a theoretical perspective, Meyer~\cite{MeyerFlexopolarization} first
proposed the $\mathrm{N_{TB}}$ and $\mathrm{N_{SB}}$ phases by linking
shape-induced spontaneous polarization to splay or bend deformations. In 1990,
we introduced a flexopolarization-induced coupling between the alignment tensor
and polarization fields that yields twist--bend order within a generalized
Landau--de~Gennes (Ginzburg--Landau--type) framework~\cite{Longa1990}, which has
since been validated quantitatively for $\mathrm{N_{TB}}$-forming CB7CB-like
mesogens~\cite{Longa2020}.

A critical advance came in 2001 with Dozov’s work~\cite{Dozov_2001}, which
generalized the Oseen--Zocher--Frank elastic theory by proposing that molecular
shapes favoring bend can reduce, and even invert, the nematic bend elastic
constant $K_{33}$, thereby stabilizing either the $\mathrm{N_{TB}}$ or
$\mathrm{N_{SB}}$ phase depending on the ratio of splay ($K_{11}$) to bend
($K_{33}$) elasticity. Specifically, if $K_{11} > 2K_{33}$, the heliconical
$\mathrm{N_{TB}}$ structure is favored, whereas if $K_{11} < 2K_{33}$, the
planar $\mathrm{N_{SB}}$ structure becomes more stable. Subsequent theoretical
work~\cite{Shamid2013statistical,Jakli2018} showed that Dozov’s elastic theory
can be recast as a flexopolarization mechanism underlying the inversion of
$K_{33}$. These studies also predicted more complex phases featuring the
coexistence of splay, bend, and twist deformations
\cite{LormanMettout2004,Shamid2014,Longa2016,Pajak2018,Rico2020,Chiappini2021,Kubala2022}
(Fig.~\ref{fig:NTB_NSB}).

Experimentally, the $\mathrm{N_{TB}}$ phase has been widely observed across
numerous thermotropic liquid-crystal systems
\cite{Jakli2018,Mandle2022,Dunmur2022}, whereas the $\mathrm{N_{SB}}$ phase and
other complex polar nematic phases remain rare, reported primarily in colloidal
systems \cite{Fernandez-Rico2020,Kotni2022} or under applied electric fields
\cite{Meyer2020,Merkel2018}. This scarcity persists despite Dozov’s relatively
broad—and, in principle, readily satisfied—elastic-constant criteria, underscoring
the need for further theoretical and computational studies to elucidate the
mechanisms governing the stability of polar nematic phases.

Motivated by these experimental findings and by the unresolved scarcity of a
stable $\mathrm{N_{SB}}$ phase—contrasting with the widespread occurrence of its
parent $\mathrm{N_{TB}}$ phase—we investigate whether confinement can stabilize
$\mathrm{N_{SB}}$. We examine the molecular organization in thin films of a
bulk-stable $\mathrm{N_{TB}}$ phase confined between parallel walls imposing
planar anchoring. Studying such confinement can clarify the mechanisms
governing the stability of splay–bend order and reveal phenomena relevant to
both fundamental research and technological applications~\cite{Panov2021}. Our
objective is to advance theoretical understanding and to inform future
experimental studies of confined bent-core molecular systems.

\begin{figure} \label{fig:molecular_model}
    \centering
    \includegraphics[width=0.6\linewidth]{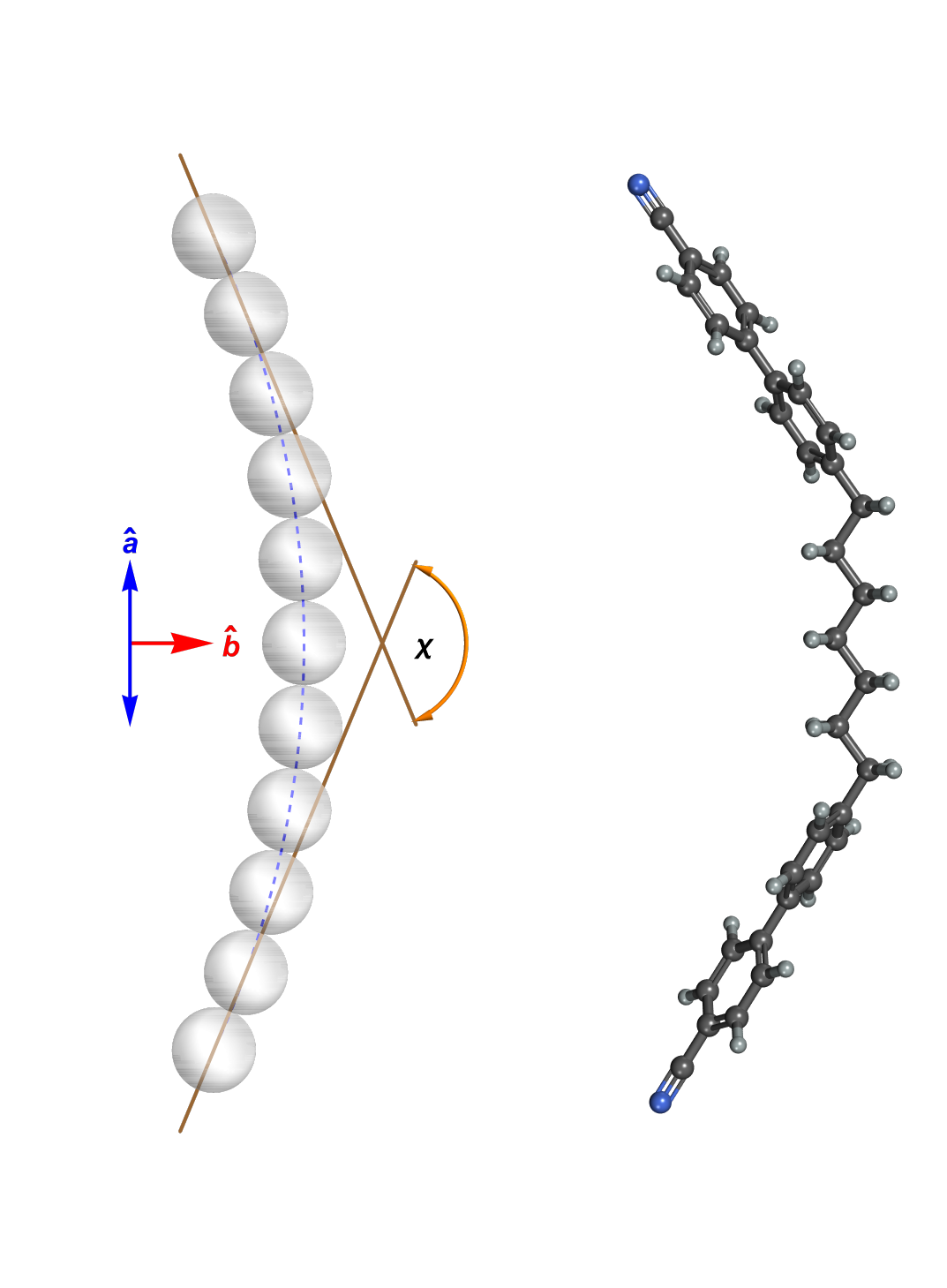}
    \caption{Model bent-shaped molecule (left) used in our simulations: a rigid
assembly of eleven identical, mutually tangent spheres arranged along a circular
arc, giving overall \(C_{2v}\) molecular symmetry. The bend angle \(\chi\) is
defined as the angle between the tangents at the terminal spheres; larger
\(\chi\) corresponds to smaller molecular curvature. The unit vectors
\(\hat{\mathbf a}\) and \(\hat{\mathbf b}\) denote the directions of the long
molecular axis and the twofold-symmetry axis, respectively. The nearest
mesogenic analogue is CB7CB (right), which exhibits a stable
\(\mathrm{N_{TB}}\) phase.}
    \label{fig:molecular_model}
\end{figure}
\begin{figure}[htb]
    \centering
    \includegraphics[width=0.8\linewidth]{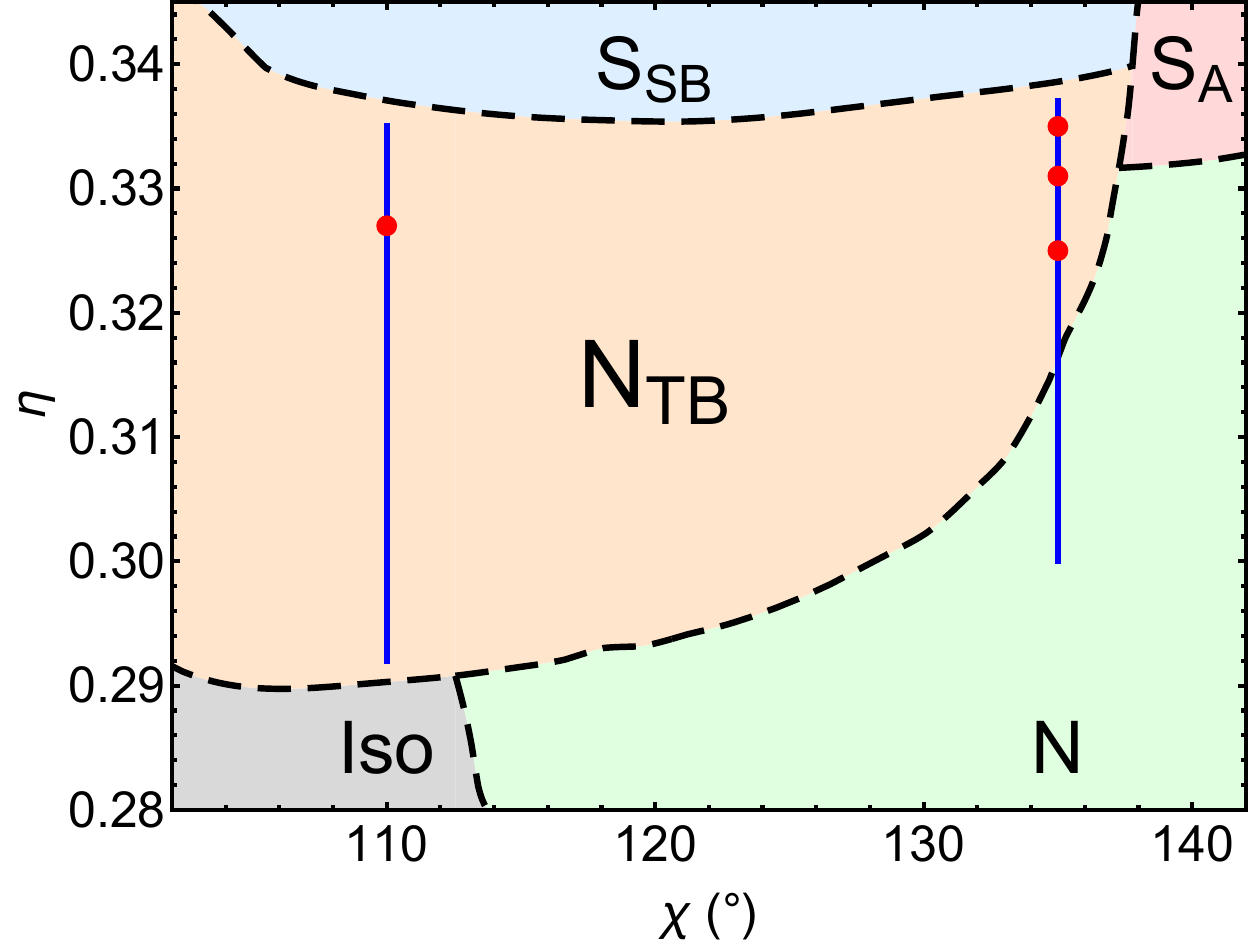}
\caption{Sketch of a partial phase diagram for the bent-core (banana-shaped) 
molecular model depicted in Fig.~(\ref{fig:phase_diagram}), shown as a function of
bend angle $\chi$ and packing fraction $\eta$. The two blue lines indicate the
simulation paths explored in this work, and the red dots mark the specific state
points for which detailed results are presented.}
\label{fig:phase_diagram}
\end{figure}

\textit{Model.---} To investigate the effects of confinement on bent-core
nematics, we employed two closely related coarse-grained models that capture
essential features of molecular ordering. In our MC simulations, performed in
the constant-pressure ensemble, each molecule was modeled as a rigid assembly of
eleven mutually tangent spheres (diameter \(\sigma = 1\)) arranged
equidistantly along a circular arc with a tunable bend angle \(\chi\) ranging
from \(180^\circ\) (linear chain) to \(0^\circ\) (semicircle)
(see Fig.~\ref{fig:molecular_model}). For MD simulations, the hard-sphere
repulsion of the MC model was replaced by the truncated and shifted repulsive
part of the Lennard–Jones potential, i.e., the differentiable
Weeks–Chandler–Andersen (WCA) interaction~\cite{WCA1971,WCA1983}. The WCA
sphere diameter was matched to its hard-sphere counterpart via the
Heyes–Okumura formula~\cite{Heyes2006}, thereby ensuring quantitative
consistency in phase behavior and observables across both models~\cite{Kubala2022}.
Greco and Ferrarini~\cite{GrecoFerrariniPRL} first showed—using MD simulations
and density-functional theory (DFT)—that packing entropy alone can stabilize the
\(\mathrm{N_{TB}}\) phase. Importantly, their molecular model was identical to
the coarse-grained arc-of-spheres model defined above. 
Kubala, Tomczyk, and Cie\'sla extended this analysis by combining MC and
MD simulations and mapping the bulk phase diagram as a function of bend angle
and packing fraction, thereby identifying the stability regions of the
\(\mathrm{N}\), \(\mathrm{N_{TB}}\), and smectic phases~\cite{Kubala2022}.
Building on these results (see Fig.~\ref{fig:phase_diagram}), our present work
focuses on the confinement-induced structural organization of the
\(\mathrm{N_{TB}}\) phase between two parallel walls.

\textit{Simulation Methods.---} To investigate the effects of confinement on
bent-core nematics, we employed two closely related coarse-grained models described above. 
MC simulations were performed
in the isothermal–isobaric (isotension) ensemble using custom software
developed by P.K. (see \textit{Code and datasets availability}). MD simulations
were carried out with LAMMPS~\cite{LAMMPS}, using both \(NpT\) and \(NVT\)
ensembles.

We considered a confinement geometry in which a monodomain of the
\(\mathrm{N_{TB}}\) phase was placed between two parallel, structureless planar
walls of finite extent. The walls were oriented parallel to the $y$–$z$ plane,
and the helical wave vector of the confined \(\mathrm{N_{TB}}\) domain was
aligned with the $z$ axis. Periodic boundary conditions were applied along $y$
and $z$ (parallel to the walls). Initial configurations were prepared by
equilibrating bulk samples (see Ref.~\cite{Kubala2022}) and then introducing the
confining walls. System sizes reached up to \(N=12{,}000\) molecules (MC) and
\(N=24{,}000\) molecules (MD). Equilibration ran up to \(3\times 10^8\) MC
cycles and \(5\times 10^6\) MD steps, followed by production runs of
\(3\times 10^8\) MC cycles and \(6\times 10^7\) MD steps for ensemble
averaging. Walls were planar and structureless; in MD simulations,
wall–particle interactions were modeled with the WCA potential. Both approaches
yielded quantitatively consistent results.

\begin{figure}
    \centering
    \includegraphics[width=0.93\linewidth]{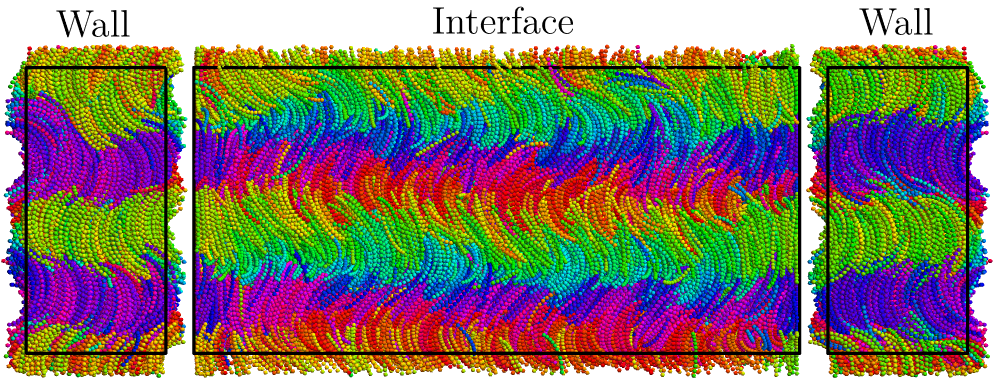}
    \includegraphics[width=1\linewidth]{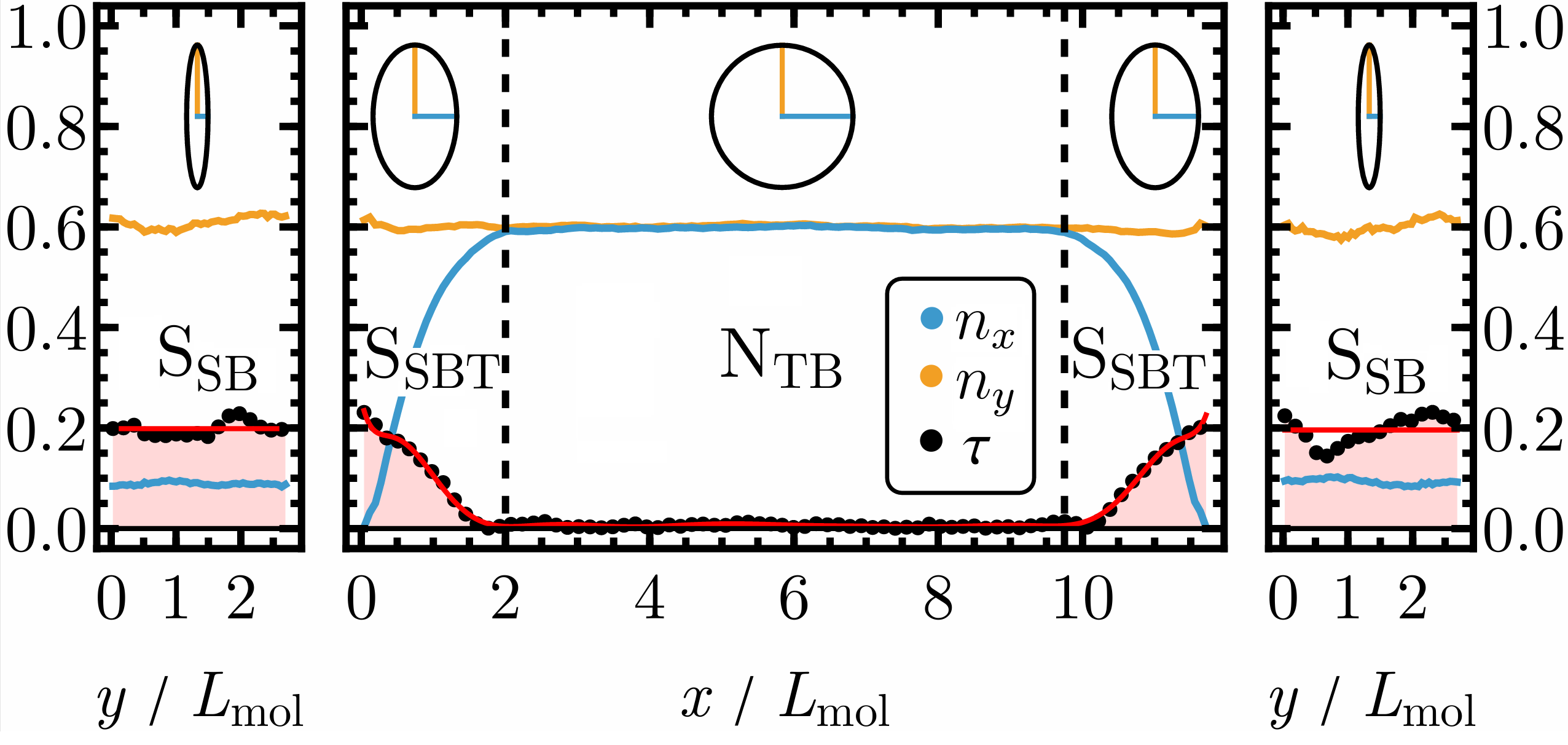}
    \caption{%
    Results of MC simulations of the \(\mathrm{N_{TB}}\) phase
    confined between two parallel walls for \(N = 12{,}000\) molecules
    with bend angle \(\chi = 110^\circ\) and packing fraction \(\eta = 0.327\). Here,
    \(L_{{mol}}\) denotes the molecular length 
    at \(\chi = 180^\circ\) (\(L_{{mol}} = 11\)).
    \textit{Top panel:} Simulation snapshot showing molecular
    organization between parallel walls. Molecular orientations are color-coded
    by the projection of the polarization axis $\mathbf{\hat{b}}$ onto the \(xy\)
    plane, perpendicular to the wave vector \(\mathbf k\).
    \textit{Bottom panel:} Smectic order parameter \(\tau\) and the 
    director projections onto the $xy$ plane as functions of distance from the
    left wall. In the \(\mathrm{N_{TB}}\) phase, the projection traces a circle;
    in the other phases, an ellipse. Sketches indicate the short and
    long semi-axes; complete ellipses are shown as insets. A sequence of three
    phases, $\mathrm{S_{SB}}$, $\mathrm{S_{SBT}}$, and $\mathrm{N_{TB}}$)—is observed
    upon moving from the wall toward the center of the sample. As the distance from 
    the wall increases, the splay component weakens and eventually vanishes on entering the
\(\mathrm{N_{TB}}\) phase.
    }
    \label{fig:MC_full_column}
\end{figure}

\begin{figure}
  \includegraphics[width=1.0\linewidth]{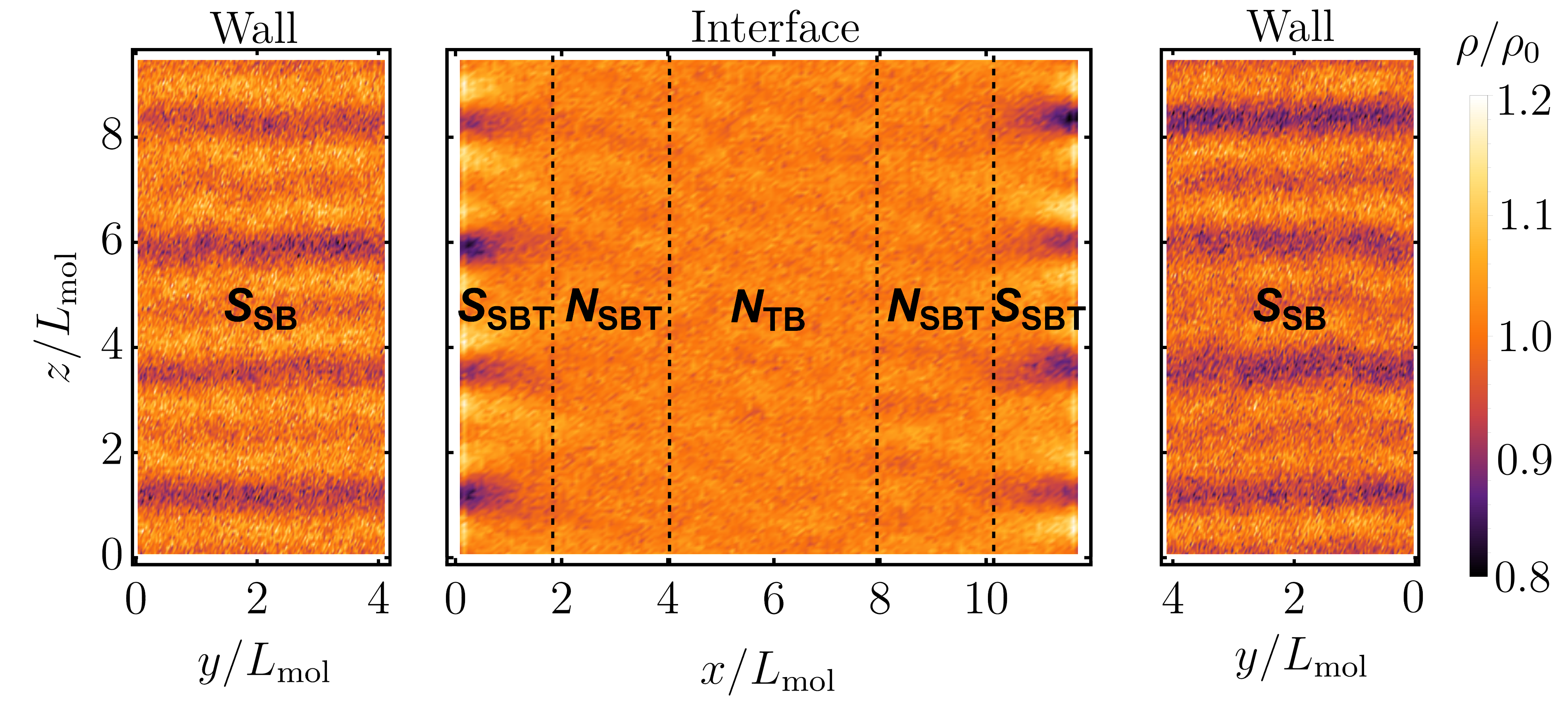}
  \centering
  \includegraphics[width=0.92\linewidth]{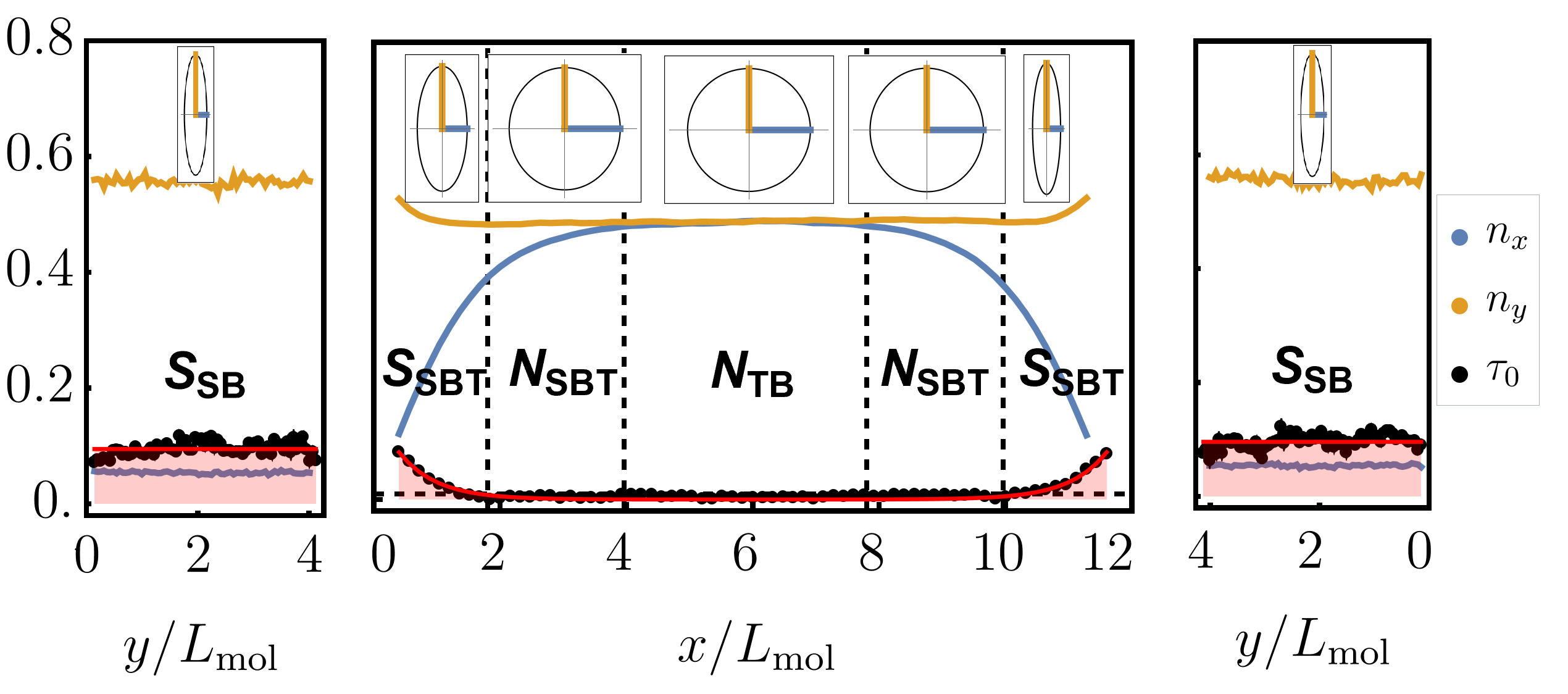}

\caption{%
Results of MD simulations of the \(\mathrm{N_{TB}}\) phase
confined between two parallel walls for \(N=24{,}000\) molecules, bend angle
\(\chi=135^\circ\), and packing fraction \(\eta=0.335\).
(Here, \(L_{\mathrm{mol}}\) is defined as in Fig.~\ref{fig:MC_full_column}.)
\textit{Top panel:} Local number-density profiles across the slit and at the
walls, normalized by the average density \(\rho_0\). Pronounced smectic layering
develops at the walls and decays approximately exponentially with distance
toward the slit center, where the bulk-stable \(\mathrm{N_{TB}}\) phase is
recovered.
\textit{Bottom panel:} Smectic order parameter \(\tau\) and the director
projection \((n_x,n_y)\) onto the \(xy\) plane as functions of distance from
the left wall. In the \(\mathrm{N_{TB}}\) phase, the locus of \((n_x,n_y)\) is a
circle; in the other phases it is an ellipse. Sketches indicate the short
\((n_x)\) and long \((n_y)\) semiaxes; complete ellipses are shown as insets.
With increasing distance from the wall, a sequence of four phases—
\(\mathrm{S_{SB}}\), \(\mathrm{S_{SBT}}\), \(\mathrm{N_{SBT}}\), and
\(\mathrm{N_{TB}}\)—is observed. The splay component weakens and ultimately
vanishes on entering the \(\mathrm{N_{TB}}\) phase.%
}
  \label{fig:MD_full_column}
\end{figure}

\begin{figure}
  \includegraphics[width=1.0\linewidth]{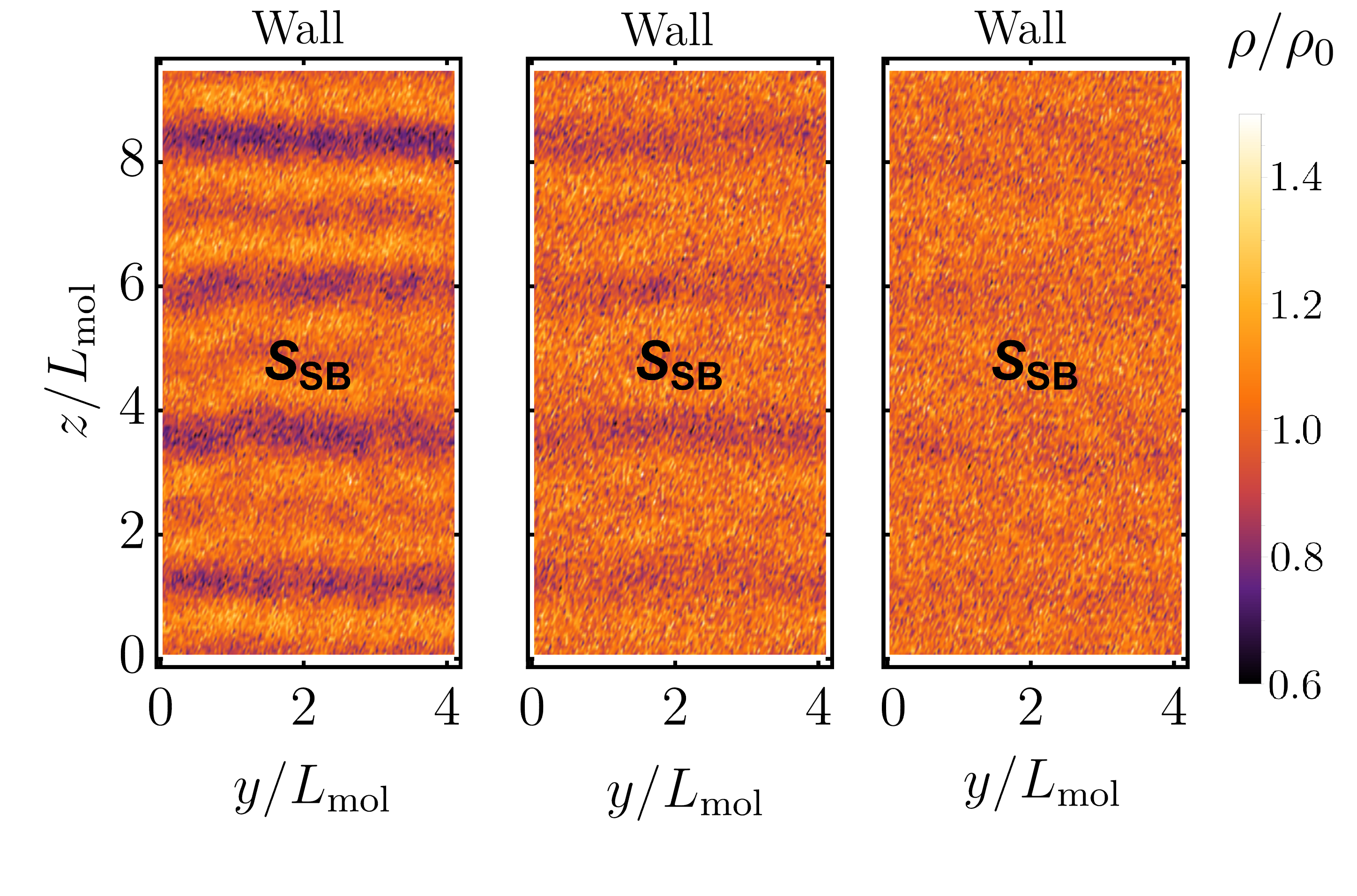}
  \includegraphics[width=0.82\linewidth]{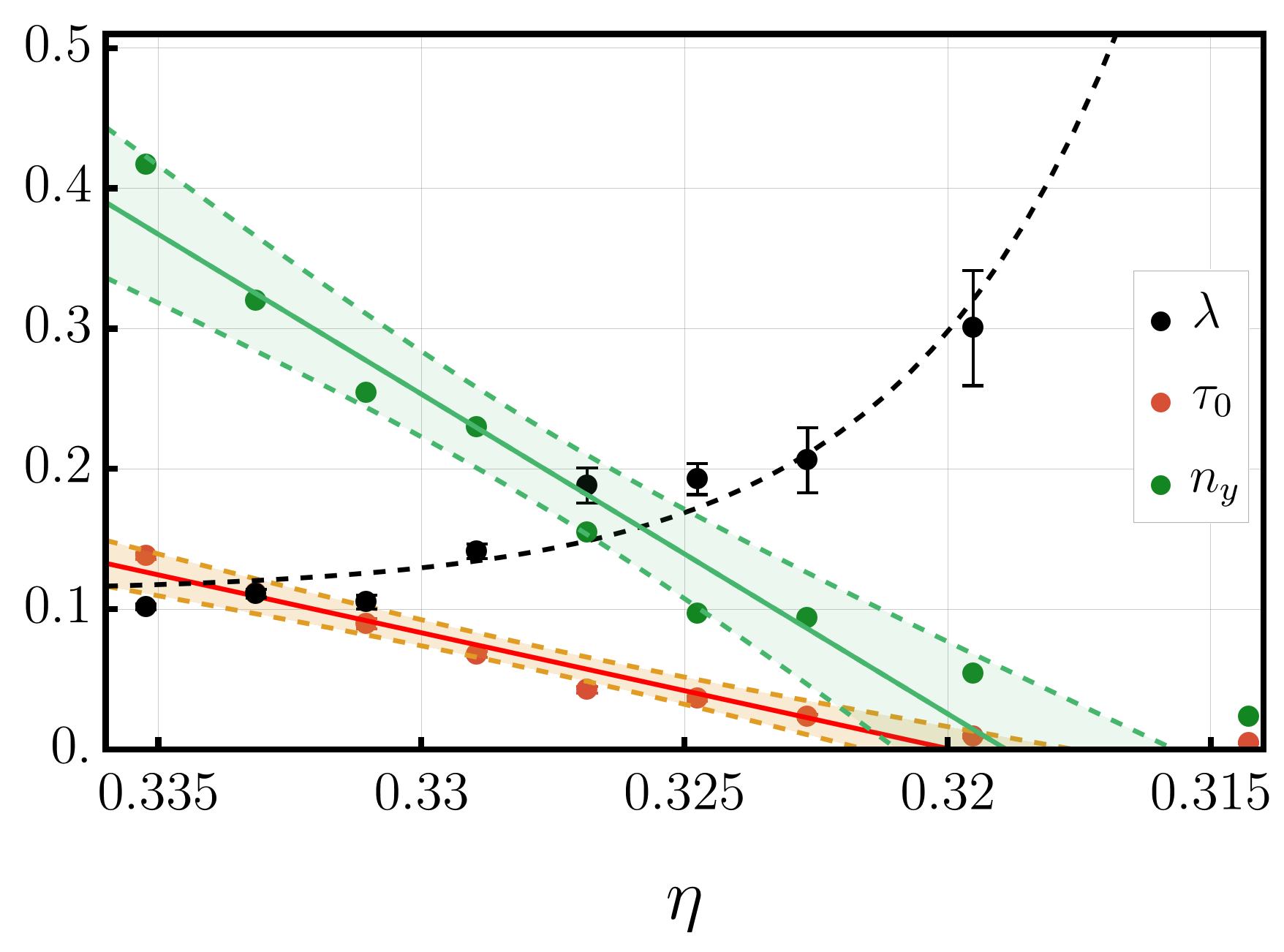}
 \caption{%
Surface ordering from molecular-dynamics (MD) simulations of the
\(\mathrm{N_{TB}}\) phase confined between two parallel walls enforcing planar
anchoring, for bend angle \(\chi=135^\circ\).
(Here, \(L_{\mathrm{mol}}\) is defined as in Fig.~\ref{fig:MC_full_column}.)
\textit{Top panel (translational order at the wall):}
Number–density maps of molecular centers at the wall for decreasing packing
fractions for \(N=24{,}000\) molecules, with the color scale normalized by the average density \(\rho_0\).
Left to right: \(\eta\in\{0.335,\,0.331,\,0.325\}\).
\textit{Bottom panel (ordering at the wall):}
Smectic order parameter \(\tau_0\) (red), correlation length \(\lambda\) (black),
and \(n_y\) (green) at the wall as functions of the packing fraction \(\eta\) for a system of \(N=6{,}000\) molecules and bend angle \(\chi=135^\circ\),
quantifying \(\mathrm{S_{SB}}\) ordering. Both the density modulation and the
splay–bend order vanish near \(\eta \approx 0.32\).}
  \label{fig:MD_full_wall}
\end{figure}

\textit{Results.---}%
Detailed simulations were carried out along the blue lines in
Fig.~\ref{fig:phase_diagram}. In all cases, the equilibrium order observed at the walls is
a smectic splay–bend (\(\mathrm{S_{SB}}\)) phase, in which director modulation
is coupled to density modulation. Moving away from the walls toward the center
of the sample—where the bulk \(\mathrm{N_{TB}}\) phase is stable—the splay
distortions and density modulations decay through a sequence of
intermediate structures. Representative results for \(\chi=110^\circ\) (MC) and
\(\chi=135^\circ\) (MD) are shown in
Figs.~\ref{fig:MC_full_column}--\ref{fig:MD_full_wall}.

Near the \(\mathrm{S_{SB}}\!-\!\mathrm{S_A}\) coexistence wedge and close to the 
\(\mathrm{N_{TB}}\!-\!\mathrm{S_{A}}\) boundary (Fig.~\ref{fig:phase_diagram}), the intermediate
structures are characterized by \(\mathrm{N_{SBT}}\) ordering adjacent to
\(\mathrm{N_{TB}}\) regions. Upon approaching the walls, the
\(\mathrm{N_{SBT}}\) phase gradually transforms into \(\mathrm{S_{SBT}}\), which
ultimately converts into \(\mathrm{S_{SB}}\) near the walls. This evolution of
the \(\mathrm{N_{TB}}\) phase is illustrated in the top panel of
Fig.~\ref{fig:MD_full_column} for \(\chi=135^\circ\) and packing fraction
\(\eta=0.335\).

Furthermore, as the packing fraction decreases, the local smectic order parameter
\(\tau\), defined as
\[
\tau(x)=\frac{1}{N T}\left|\sum_{t=1}^{T}\sum_{n=1}^{N}
\Theta\!\big(\Delta-|x_n(t)-x|\big)\,
e^{2\pi i k\, z_n(t)/Z}\right|,
\]
also decreases (Fig.~\ref{fig:MD_full_wall}). Here, the sums run over \(N\)
molecules and \(T\) configuration snapshots; \(z_n(t)\), \(Z\), and \(k\) denote
the \(z\)-position of the \(n\)-th molecule at time \(t\), the box length along
\(z\), and the number of density–modulation periods along the wave–vector
direction, respectively. The Heaviside step function \(\Theta\) restricts the
average to molecules whose centers lie within a slab of half-width
\(\Delta=0.12\,L_{\mathrm{mol}}\) centered a distance \(x\) from the nearest
wall. For each \(\eta\), the resulting \(\tau(x)\) was fitted with
\(\tau(x)=\tau_0 e^{-x/\lambda}\). The \(\eta\)-dependence of \(\tau_0\) and
\(\lambda\) is shown in the bottom panel of Fig.~\ref{fig:MD_full_wall}.
With decreasing packing fraction, the \(\mathrm{S_{SB}}\) phase at the walls
also weakens and eventually disappears as \(\eta\) approaches values
characteristic of the bulk nematic or isotropic phases. 

Guided by the structural analysis above, we further quantify the interfacial
fine structure of orientational order  by decomposing the
director–gradient field into the canonical Oseen–Frank modes (splay, twist,
bend, and saddle–splay). This mode-resolved perspective provides, to our
knowledge, the first direct bridge between particle-resolved simulations of
confined \(\mathrm{N_{TB}}\) and continuum elasticity, and it pinpoints where boundaries
select distinct elastic responses—most notably how interfacial layers
accommodate chirality and activate the saddle–splay channel. Experiments and
modeling by Xia {\it{et al.}} demonstrate that suitably programmed surfaces can
control symmetry via this channel~\cite{XiaNatCommun2019}. In our system, the
channel is activated differently: competition between the heliconical bulk
\(\mathrm{N_{TB}}\) texture and planar surface anchoring selects the observed sequence of
interfacial ordering. Following Selinger’s geometric formulation, we monitor the
saddle–splay interfacial {density} on the same footing as splay, twist,
and bend~\cite{SelingerLCR2018}.

For completeness, and to connect with our maps, for \(i\in\{x,y,z\}\) we define
\(
\mathrm{splay}_i = n_i\,(\nabla\!\cdot\!\mathbf n)\) and
\(\mathrm{bend}_i = \big(\mathbf n\times(\nabla\times\mathbf n)\big)_i \),
while the pseudoscalar and scalar fields are 
\(\mathrm{twist} = \mathbf n\!\cdot\!(\nabla\times\mathbf n)\)
and
\(
\mathrm{saddle\text{–}splay} = -\,\nabla\!\cdot\!\Big(\mathbf n\,(\nabla\!\cdot\!\mathbf n)
+\mathbf n\times(\nabla\times\mathbf n)\Big)
\), respectively. 
We also evaluate the coarse-grained polarization vector field \(\mathbf p(x,z)\), which, in 
our sterically driven model, 
is given by the local average of the molecular short axis \({\hat{\mathbf{b}}}\).

\begin{figure}
  \raggedright
  \includegraphics[width=1.0\linewidth]{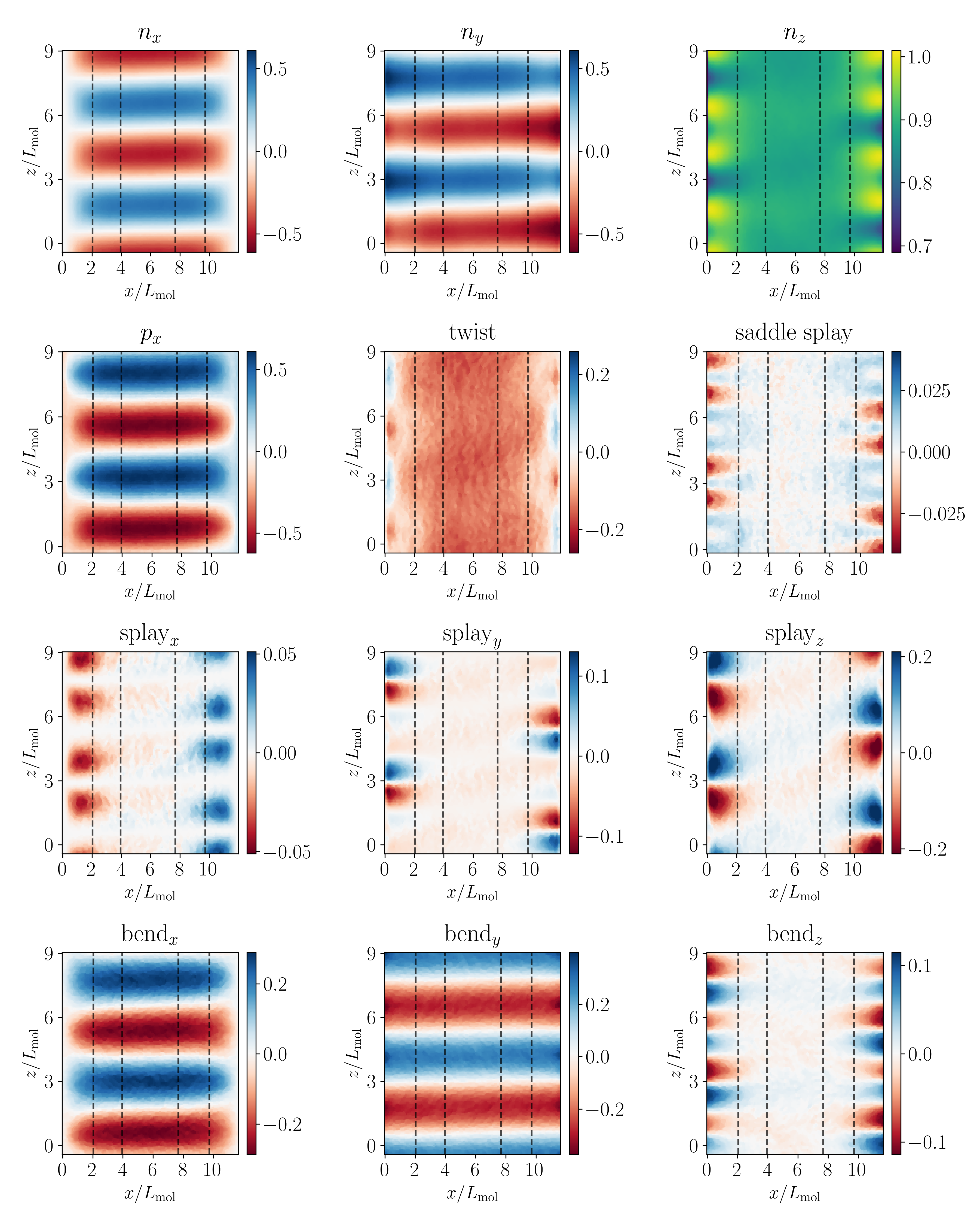}
\caption{Representative local distributions of the director, 
director–distortion modes and orientational order parameter
  between planar walls, corresponding to the two blue simulation paths in
  Fig.~\ref{fig:phase_diagram}, obtained from MD simulations
  (\(N=24{,}000\), \(\chi=135^\circ\), \(\eta=0.335\)). Top row: components of the director field. 
  Second row: the $x$ component of the polarization field, twist
  and saddle–splay. Third row: splay
  components \(\mathrm{splay}_x\), \(\mathrm{splay}_y\),
  \(\mathrm{splay}_z\). Bottom row: bend components \(\mathrm{bend}_x\),
  \(\mathrm{bend}_y\), \(\mathrm{bend}_z\).
  Axes: horizontal—\(x/L_{\mathrm{mol}}\) (distance from the left wall);
  vertical—\(z/L_{\mathrm{mol}}\) (coordinate along the bulk wave vector).
  Color scales for each panel are shown on the right. 
  The vertical dashed lines mark the phase  boundaries indicated 
in Fig.~\ref{fig:MD_full_column}.}
  \label{fig:elastic_deformations}
\end{figure}

To compute any coarse-grained observable \(A(x,z)\) from a microscopic
quantity \(A_n\), we use the same slab averaging employed for \(\tau(x)\):
\begin{equation}
\begin{split}
A(x,z)=\frac{1}{N T}\sum_{t=1}^{T}\sum_{n=1}^{N}
A_n(t)
\Theta\!\big(\Delta_x-|x_n(t)-x|\big)\,)\\[2pt] \times
\Theta\!\big(\Delta_z-|z_n(t)-z|\big),
\end{split} \nonumber
\end{equation}
where \(\Delta_x = 0.04\,L_{\mathrm{mol}}\) and
\(\Delta_z = 0.03\,L_{\mathrm{mol}}\) are the slab half–widths.
Choosing
\(A_n=\hat{\mathbf b}_n\) yields the polarization \(\mathbf p(x,z)\). Choosing
\(A_n=\tfrac{3}{2}\,\hat{\mathbf a}_n(t)\otimes\hat{\mathbf a}_n(t)
-\tfrac{1}{2}\,\mathbf I\) yields the alignment tensor
\(\mathbf Q(x,z)\). The local director \(\mathbf n(x,z)\) is then defined as
the normalized eigenvector of \(\mathbf Q(x,z)\) corresponding to its
largest-magnitude (nondegenerate) eigenvalue. Spatial derivatives are obtained
by convolving the discretized director field with standard \(3\times3\) Sobel
kernels to approximate first-order gradients~\cite{Sobel1968,GonzalezWoods2006}.
Results are shown in Fig.~\ref{fig:elastic_deformations}.

Guided by Fig.~\ref{fig:elastic_deformations}, we find a robust interfacial
orientational pattern representative of the two blue simulation paths in
Fig.~\ref{fig:phase_diagram}. Near the center of each interfacial layer, the
texture locks into a heliconical nematic twist–bend state 
(\( \mathrm{pitch}=\frac{2 \pi}{k}\approx 4.5 L_{mol}\)
in Fig.~\ref{fig:elastic_deformations}), 
and the deformation maps show the corresponding position–independent signatures. 
Along this mid–plane of the \(\mathrm{N_{TB}}\) slab
we find (up to numerical accuracy) \(\mathrm{splay}=0\), \(\mathrm{bend}_z=0\),
and saddle–splay \(=0\), while twist retains a fixed sign across the slab. 
The in–plane bend components \(\mathrm{bend}_x\) and \(\mathrm{bend}_y\)
exhibit the same periodic modulation with the expected quarter–period
phase shift along \(z\).
The  \(n_z\) component of the director is nearly
constant and less than unity, indicating saturated tilt, while 
the polarization field is essentially
collinear with bend: \(p_x\), \(p_y\), and \(p_z\) stripes with the same
wavelength and phase as \(\mathrm{bend}_x\), \(\mathrm{bend}_y\), and
\(\mathrm{bend}_z\), respectively. This is consistent with the bend–polarization
relation \(\mathbf p \sim \,\mathbf n \times (\nabla \times \mathbf n)\),
up to an overall scale and a sign set by the handedness.
These features again confirm
that the layer’s \(\mathrm{N_{TB}}\) interior is consistent with bulk twist–bend ordering and acts
as a phase–matching sub-layer between the two walls.

The boundary–driven structure—set by the competition between planar anchoring
and the heliconical bulk—is confined to the near–wall regions
\(x/L_{\max}\!\lesssim\!2\) and \(x/L_{\max}\!\gtrsim\!10\). There, \(n_z\)
displays the same axial wavelength and comparable amplitude at both walls, but
the bright/dark bands are offset by almost half a period along \(z\). This
near–antiphase relation is the signature of the improper symmetry that relates
the two interfacial skins (reflection about the mid–plane combined with a
half–pitch translation along \(z\)).
Small even–harmonic content in the boundary layers explains the slight,
systematic misalignment of extrema (crests do not map exactly onto troughs).

Across the \(S_{SBT}\) skins (splay–bend–twist), a weak, alternating twist
localizes and coexists with alternating splay and saddle–splay; this is the
entropic/elastic cost of steering the texture away from the purely planar bend
favored by the walls. Directly at the walls (\(S_{SB}\)) the twist channel is
suppressed, the splay components strengthen, and the saddle–splay shows
sign–selective lobes that are phase–locked to the splay/bend bands—consistent
with a saddle–like (negative Gaussian curvature) distortion producing surface
torques that reinforce the interfacial splay–bend texture. While splay and
saddle–splay are strongest in the smectic regions, they remain weak but finite
in the \( \mathrm{N_{SBT}} \) bands. Finally, the
\(y\)-component of the bend field is nearly identical across all regions.

Concerning polarization at the walls, \(p_x\) develops a finite mean component
with opposite signs on the two sides—numerically
\(p_{x, L}\!\approx\!-0.17\) and \(p_{x, R}\!\approx\!+0.17\).
This antisymmetric offset is consistent with the flexo–splay contribution,
proportional to \(\mathbf n\,(\nabla\!\cdot\!\mathbf n)\): the near–wall splay
\((\nabla\!\cdot\!\mathbf n)\) is odd under the mid–plane reflection
\(x\!\to\!L-x\), so the \(x\)-projection of \(\mathbf p\) changes sign from one
wall to the other. By contrast, \(\mathrm{bend}_x\) at the walls is dominated by
its oscillatory fundamental and has (nearly) zero mean, so it carries no
comparable constant component. Consequently, the sample–averaged polarization
along \(x\) vanishes by symmetry, while the two walls host equal–and–opposite
interfacial polarizations.

\textit{Discussion.---}
Understanding how periodically modulated {\it{polar}} nematics form and remain
stable is pivotal for advancing liquid–crystal theory and enabling
reconfigurable optical elements and display concepts.
With planar anchoring and the wavevector parallel to the plates, confinement
selects a robust wall–to–wall architecture: thin \(\mathrm{S_{SB}}\) skins at
the walls (as in strictly 2D flexible bend–core systems
\cite{Longa2023}), \(\mathrm{S_{SBT}}\) buffers, and—near the
\(\mathrm{N_{TB}}\)–smectic threshold—an interior \(\mathrm{N_{SBT}}\) band,
phase–matched by a heliconical \(\mathrm{N_{TB}}\) core. Mode–resolved maps
show  families of distortions along \(z\), strong splay
and saddle–splay at boundaries, a fixed–sign twist across the slab, and a
half–pitch antiphase of \(n_z\) between the walls. The polarization
\(\mathbf p\) forms stripes in phase with bend in the \(\mathrm{N_{TB}}\)
interior, while equal–and–opposite mean \(p_x\) develops at the two walls, so
the sample–averaged polarization vanishes.

Across our representative sweep of the phase diagram we did {not} find a
stable \(\mathrm{N_{SB}}\) in three dimensions; to our knowledge it has only
been stabilized in strictly 2D boomerang models
\cite{Karbowniczek2017}. This rationalizes the preference for twist-mediated
\(\mathrm{S_{SBT}}\)–\(\mathrm{N_{SBT}}\) pathways over a pure
\(\mathrm{N_{SB}}\) phase. Mechanistically, the saddle–splay {density}
concentrates at walls and at \(\mathrm{S_{SB}}\!\leftrightarrow\!\mathrm{S_{SBT}}\)
crossovers; this boundary channel allows confinement to “program” symmetry and
handedness, consistent with the surface–driven control reported by
\cite{XiaNatCommun2019}. Our results, bridging particle simulations of confined
\(\mathrm{N_{TB}}\) with continuum elasticity, provide practical design rules:
by tuning anchoring, geometry, and proximity to the \(\mathrm{N_{TB}}\)–smectic
threshold, one can assemble prescribed
\(\mathrm{S_{SB}}/\mathrm{S_{SBT}}/\mathrm{N_{SBT}}/\mathrm{N_{TB}}\) stacks and
program interfacial chirality and polarization for chiral photonics,
polarization gratings, and low–power electro–optic devices.


\textit{Code and datasets availability---}
The source code of an original RAMPACK simulation package used to perform
Monte Carlo sampling is available at \url{https://github.com/PKua007/rampack}. 
The input script for LAMMPS and RAMPACK along with the datasets generated 
during and/or analyzed during the current
study are available from 
S.D. and P.K. upon reasonable request. \\

\acknowledgements

\textit{Acknowledgements---}
We thank Agnieszka Chrzanowska 
and Michal Cie\'sla for insightful comments at various stages of this work. 
L.L. is grateful to Mark Dennis
for helpful suggestions regarding the calculation of the saddle-splay contribution.
The authors acknowledge the support of the National Science 
Centre in Poland grant no. 2021/43/B/ST3/03135. Numerical simulations were carried out
with the support of the Interdisciplinary Center for Mathematical and Computational
Modeling (ICM) at the University of Warsaw under grant no. G27-8.

\bibliography{main}

\begin{thebibliography}{42}%
\makeatletter
\providecommand \@ifxundefined [1]{%
 \@ifx{#1\undefined}
}%
\providecommand \@ifnum [1]{%
 \ifnum #1\expandafter \@firstoftwo
 \else \expandafter \@secondoftwo
 \fi
}%
\providecommand \@ifx [1]{%
 \ifx #1\expandafter \@firstoftwo
 \else \expandafter \@secondoftwo
 \fi
}%
\providecommand \natexlab [1]{#1}%
\providecommand \enquote  [1]{``#1''}%
\providecommand \bibnamefont  [1]{#1}%
\providecommand \bibfnamefont [1]{#1}%
\providecommand \citenamefont [1]{#1}%
\providecommand \href@noop [0]{\@secondoftwo}%
\providecommand \href [0]{\begingroup \@sanitize@url \@href}%
\providecommand \@href[1]{\@@startlink{#1}\@@href}%
\providecommand \@@href[1]{\endgroup#1\@@endlink}%
\providecommand \@sanitize@url [0]{\catcode `\\12\catcode `\$12\catcode `\&12\catcode `\#12\catcode `\^12\catcode `\_12\catcode `\%12\relax}%
\providecommand \@@startlink[1]{}%
\providecommand \@@endlink[0]{}%
\providecommand \url  [0]{\begingroup\@sanitize@url \@url }%
\providecommand \@url [1]{\endgroup\@href {#1}{\urlprefix }}%
\providecommand \urlprefix  [0]{URL }%
\providecommand \Eprint [0]{\href }%
\providecommand \doibase [0]{https://doi.org/}%
\providecommand \selectlanguage [0]{\@gobble}%
\providecommand \bibinfo  [0]{\@secondoftwo}%
\providecommand \bibfield  [0]{\@secondoftwo}%
\providecommand \translation [1]{[#1]}%
\providecommand \BibitemOpen [0]{}%
\providecommand \bibitemStop [0]{}%
\providecommand \bibitemNoStop [0]{.\EOS\space}%
\providecommand \EOS [0]{\spacefactor3000\relax}%
\providecommand \BibitemShut  [1]{\csname bibitem#1\endcsname}%
\let\auto@bib@innerbib\@empty
\bibitem [{\citenamefont {Cestari}\ \emph {et~al.}(2011)\citenamefont {Cestari}, \citenamefont {Diez-Berart}, \citenamefont {Dunmur}, \citenamefont {Ferrarini}, \citenamefont {de~la Fuente}, \citenamefont {Jackson}, \citenamefont {Lopez}, \citenamefont {Luckhurst}, \citenamefont {Perez-Jubindo}, \citenamefont {Richardson}, \citenamefont {Salud}, \citenamefont {Timimi},\ and\ \citenamefont {Zimmermann}}]{Cestari2011}%
  \BibitemOpen
  \bibfield  {author} {\bibinfo {author} {\bibfnamefont {M.}~\bibnamefont {Cestari}}, \bibinfo {author} {\bibfnamefont {S.}~\bibnamefont {Diez-Berart}}, \bibinfo {author} {\bibfnamefont {D.~A.}\ \bibnamefont {Dunmur}}, \bibinfo {author} {\bibfnamefont {A.}~\bibnamefont {Ferrarini}}, \bibinfo {author} {\bibfnamefont {M.~R.}\ \bibnamefont {de~la Fuente}}, \bibinfo {author} {\bibfnamefont {D.~J.~B.}\ \bibnamefont {Jackson}}, \bibinfo {author} {\bibfnamefont {D.~O.}\ \bibnamefont {Lopez}}, \bibinfo {author} {\bibfnamefont {G.~R.}\ \bibnamefont {Luckhurst}}, \bibinfo {author} {\bibfnamefont {M.~A.}\ \bibnamefont {Perez-Jubindo}}, \bibinfo {author} {\bibfnamefont {R.~M.}\ \bibnamefont {Richardson}}, \bibinfo {author} {\bibfnamefont {J.}~\bibnamefont {Salud}}, \bibinfo {author} {\bibfnamefont {B.~A.}\ \bibnamefont {Timimi}},\ and\ \bibinfo {author} {\bibfnamefont {H.}~\bibnamefont {Zimmermann}},\ }\bibfield  {title} {\bibinfo {title} {Phase behavior and properties of the liquid-crystal dimer
  1\ensuremath{'}\ensuremath{'},7\ensuremath{'}\ensuremath{'}-bis(4-cyanobiphenyl-4\ensuremath{'}-yl) heptane: A twist-bend nematic liquid crystal},\ }\href {https://doi.org/10.1103/PhysRevE.84.031704} {\bibfield  {journal} {\bibinfo  {journal} {Phys. Rev. E}\ }\textbf {\bibinfo {volume} {84}},\ \bibinfo {pages} {031704} (\bibinfo {year} {2011})}\BibitemShut {NoStop}%
\bibitem [{\citenamefont {Borshch}\ \emph {et~al.}(2013)\citenamefont {Borshch}, \citenamefont {Kim}, \citenamefont {Xiang}, \citenamefont {Gao}, \citenamefont {J{\'a}kli}, \citenamefont {Panov}, \citenamefont {Vij}, \citenamefont {Imrie}, \citenamefont {Tamba}, \citenamefont {Mehl},\ and\ \citenamefont {Lavrentovich}}]{Borshch2013}%
  \BibitemOpen
  \bibfield  {author} {\bibinfo {author} {\bibfnamefont {V.}~\bibnamefont {Borshch}}, \bibinfo {author} {\bibfnamefont {Y.-K.}\ \bibnamefont {Kim}}, \bibinfo {author} {\bibfnamefont {J.}~\bibnamefont {Xiang}}, \bibinfo {author} {\bibfnamefont {M.}~\bibnamefont {Gao}}, \bibinfo {author} {\bibfnamefont {A.}~\bibnamefont {J{\'a}kli}}, \bibinfo {author} {\bibfnamefont {V.~P.}\ \bibnamefont {Panov}}, \bibinfo {author} {\bibfnamefont {J.~K.}\ \bibnamefont {Vij}}, \bibinfo {author} {\bibfnamefont {C.~T.}\ \bibnamefont {Imrie}}, \bibinfo {author} {\bibfnamefont {M.~G.}\ \bibnamefont {Tamba}}, \bibinfo {author} {\bibfnamefont {G.~H.}\ \bibnamefont {Mehl}},\ and\ \bibinfo {author} {\bibfnamefont {O.~D.}\ \bibnamefont {Lavrentovich}},\ }\bibfield  {title} {\bibinfo {title} {Nematic twist-bend phase with nanoscale modulation of molecular orientation},\ }\href {https://doi.org/10.1038/ncomms3635} {\bibfield  {journal} {\bibinfo  {journal} {Nat. Commun.}\ }\textbf {\bibinfo {volume} {4}},\ \bibinfo {pages} {2635} (\bibinfo
  {year} {2013})}\BibitemShut {NoStop}%
\bibitem [{\citenamefont {Chen}\ \emph {et~al.}(2013)\citenamefont {Chen}, \citenamefont {Porada}, \citenamefont {Hooper}, \citenamefont {Klittnick}, \citenamefont {Shen}, \citenamefont {Tuchband}, \citenamefont {Korblova}, \citenamefont {Bedrov}, \citenamefont {Walba}, \citenamefont {Glaser}, \citenamefont {Maclennan},\ and\ \citenamefont {Clark}}]{Chen2013}%
  \BibitemOpen
  \bibfield  {author} {\bibinfo {author} {\bibfnamefont {D.}~\bibnamefont {Chen}}, \bibinfo {author} {\bibfnamefont {J.~H.}\ \bibnamefont {Porada}}, \bibinfo {author} {\bibfnamefont {J.~B.}\ \bibnamefont {Hooper}}, \bibinfo {author} {\bibfnamefont {A.}~\bibnamefont {Klittnick}}, \bibinfo {author} {\bibfnamefont {Y.}~\bibnamefont {Shen}}, \bibinfo {author} {\bibfnamefont {M.~R.}\ \bibnamefont {Tuchband}}, \bibinfo {author} {\bibfnamefont {E.}~\bibnamefont {Korblova}}, \bibinfo {author} {\bibfnamefont {D.}~\bibnamefont {Bedrov}}, \bibinfo {author} {\bibfnamefont {D.~M.}\ \bibnamefont {Walba}}, \bibinfo {author} {\bibfnamefont {M.~A.}\ \bibnamefont {Glaser}}, \bibinfo {author} {\bibfnamefont {J.~E.}\ \bibnamefont {Maclennan}},\ and\ \bibinfo {author} {\bibfnamefont {N.~A.}\ \bibnamefont {Clark}},\ }\bibfield  {title} {\bibinfo {title} {Chiral heliconical ground state of nanoscale pitch in a nematic liquid crystal of achiral molecular dimers},\ }\href {https://doi.org/10.1073/pnas.1314654110} {\bibfield
  {journal} {\bibinfo  {journal} {Proc. Natl. Acad. Sci. U.S.A.}\ }\textbf {\bibinfo {volume} {110}},\ \bibinfo {pages} {15931} (\bibinfo {year} {2013})}\BibitemShut {NoStop}%
\bibitem [{\citenamefont {Jákli}\ \emph {et~al.}(2018)\citenamefont {Jákli}, \citenamefont {Lavrentovich},\ and\ \citenamefont {Selinger}}]{Jakli2018}%
  \BibitemOpen
  \bibfield  {author} {\bibinfo {author} {\bibfnamefont {A.}~\bibnamefont {Jákli}}, \bibinfo {author} {\bibfnamefont {O.~D.}\ \bibnamefont {Lavrentovich}},\ and\ \bibinfo {author} {\bibfnamefont {J.~V.}\ \bibnamefont {Selinger}},\ }\bibfield  {title} {\bibinfo {title} {Physics of liquid crystals of bent-shaped molecules},\ }\href {https://doi.org/10.1103/RevModPhys.90.045004} {\bibfield  {journal} {\bibinfo  {journal} {Reviews of Modern Physics}\ }\textbf {\bibinfo {volume} {90}},\ \bibinfo {pages} {045004} (\bibinfo {year} {2018})}\BibitemShut {NoStop}%
\bibitem [{\citenamefont {Mertelj}\ \emph {et~al.}(2018{\natexlab{a}})\citenamefont {Mertelj}, \citenamefont {Cmok}, \citenamefont {Sebasti\'an}, \citenamefont {Mandle}, \citenamefont {Parker}, \citenamefont {Whitwood}, \citenamefont {Goodby},\ and\ \citenamefont {\ifmmode \check{C}\else \v{C}\fi{}opi\ifmmode~\check{c}\else \v{c}\fi{}}}]{PhysRevX.8.041025}%
  \BibitemOpen
  \bibfield  {author} {\bibinfo {author} {\bibfnamefont {A.}~\bibnamefont {Mertelj}}, \bibinfo {author} {\bibfnamefont {L.}~\bibnamefont {Cmok}}, \bibinfo {author} {\bibfnamefont {N.}~\bibnamefont {Sebasti\'an}}, \bibinfo {author} {\bibfnamefont {R.~J.}\ \bibnamefont {Mandle}}, \bibinfo {author} {\bibfnamefont {R.~R.}\ \bibnamefont {Parker}}, \bibinfo {author} {\bibfnamefont {A.~C.}\ \bibnamefont {Whitwood}}, \bibinfo {author} {\bibfnamefont {J.~W.}\ \bibnamefont {Goodby}},\ and\ \bibinfo {author} {\bibfnamefont {M.}~\bibnamefont {\ifmmode \check{C}\else \v{C}\fi{}opi\ifmmode~\check{c}\else \v{c}\fi{}}},\ }\bibfield  {title} {\bibinfo {title} {Splay nematic phase},\ }\href {https://doi.org/10.1103/PhysRevX.8.041025} {\bibfield  {journal} {\bibinfo  {journal} {Phys. Rev. X}\ }\textbf {\bibinfo {volume} {8}},\ \bibinfo {pages} {041025} (\bibinfo {year} {2018}{\natexlab{a}})}\BibitemShut {NoStop}%
\bibitem [{\citenamefont {Sebasti\'an}\ \emph {et~al.}(2020)\citenamefont {Sebasti\'an}, \citenamefont {Cmok}, \citenamefont {Mandle}, \citenamefont {de~la Fuente}, \citenamefont {Dreven\ifmmode \check{s}\else~\v{s}\fi{}ek Olenik}, \citenamefont {\ifmmode \check{C}\else \v{C}\fi{}opi\ifmmode~\check{c}\else \v{c}\fi{}},\ and\ \citenamefont {Mertelj}}]{PhysRevLett.124.037801}%
  \BibitemOpen
  \bibfield  {author} {\bibinfo {author} {\bibfnamefont {N.}~\bibnamefont {Sebasti\'an}}, \bibinfo {author} {\bibfnamefont {L.}~\bibnamefont {Cmok}}, \bibinfo {author} {\bibfnamefont {R.~J.}\ \bibnamefont {Mandle}}, \bibinfo {author} {\bibfnamefont {M.~R.}\ \bibnamefont {de~la Fuente}}, \bibinfo {author} {\bibfnamefont {I.}~\bibnamefont {Dreven\ifmmode \check{s}\else~\v{s}\fi{}ek Olenik}}, \bibinfo {author} {\bibfnamefont {M.}~\bibnamefont {\ifmmode \check{C}\else \v{C}\fi{}opi\ifmmode~\check{c}\else \v{c}\fi{}}},\ and\ \bibinfo {author} {\bibfnamefont {A.}~\bibnamefont {Mertelj}},\ }\bibfield  {title} {\bibinfo {title} {Ferroelectric-ferroelastic phase transition in a nematic liquid crystal},\ }\href {https://doi.org/10.1103/PhysRevLett.124.037801} {\bibfield  {journal} {\bibinfo  {journal} {Phys. Rev. Let.}\ }\textbf {\bibinfo {volume} {124}},\ \bibinfo {pages} {037801} (\bibinfo {year} {2020})}\BibitemShut {NoStop}%
\bibitem [{\citenamefont {Chen}\ \emph {et~al.}(2020{\natexlab{a}})\citenamefont {Chen}, \citenamefont {Korblova}, \citenamefont {Dong}, \citenamefont {Wei}, \citenamefont {Shao}, \citenamefont {Radzihovsky}, \citenamefont {Glaser}, \citenamefont {Maclennan}, \citenamefont {Bedrov}, \citenamefont {Walba} \emph {et~al.}}]{chen2020first}%
  \BibitemOpen
  \bibfield  {author} {\bibinfo {author} {\bibfnamefont {X.}~\bibnamefont {Chen}}, \bibinfo {author} {\bibfnamefont {E.}~\bibnamefont {Korblova}}, \bibinfo {author} {\bibfnamefont {D.}~\bibnamefont {Dong}}, \bibinfo {author} {\bibfnamefont {X.}~\bibnamefont {Wei}}, \bibinfo {author} {\bibfnamefont {R.}~\bibnamefont {Shao}}, \bibinfo {author} {\bibfnamefont {L.}~\bibnamefont {Radzihovsky}}, \bibinfo {author} {\bibfnamefont {M.~A.}\ \bibnamefont {Glaser}}, \bibinfo {author} {\bibfnamefont {J.~E.}\ \bibnamefont {Maclennan}}, \bibinfo {author} {\bibfnamefont {D.}~\bibnamefont {Bedrov}}, \bibinfo {author} {\bibfnamefont {D.~M.}\ \bibnamefont {Walba}}, \emph {et~al.},\ }\bibfield  {title} {\bibinfo {title} {First-principles experimental demonstration of ferroelectricity in a thermotropic nematic liquid crystal: Polar domains and striking electro-optics},\ }\href@noop {} {\bibfield  {journal} {\bibinfo  {journal} {Proc. Nat. Acad. Sci.}\ }\textbf {\bibinfo {volume} {117}},\ \bibinfo {pages} {14021} (\bibinfo {year}
  {2020}{\natexlab{a}})}\BibitemShut {NoStop}%
\bibitem [{\citenamefont {Mandle}\ \emph {et~al.}(2021)\citenamefont {Mandle}, \citenamefont {Sebasti{\'a}n}, \citenamefont {Martinez-Perdiguero},\ and\ \citenamefont {Mertelj}}]{Mandle2021}%
  \BibitemOpen
  \bibfield  {author} {\bibinfo {author} {\bibfnamefont {R.~J.}\ \bibnamefont {Mandle}}, \bibinfo {author} {\bibfnamefont {N.}~\bibnamefont {Sebasti{\'a}n}}, \bibinfo {author} {\bibfnamefont {J.}~\bibnamefont {Martinez-Perdiguero}},\ and\ \bibinfo {author} {\bibfnamefont {A.}~\bibnamefont {Mertelj}},\ }\bibfield  {title} {\bibinfo {title} {On the molecular origins of the ferroelectric splay nematic phase},\ }\href@noop {} {\bibfield  {journal} {\bibinfo  {journal} {Nat. commun.}\ }\textbf {\bibinfo {volume} {12}},\ \bibinfo {pages} {1} (\bibinfo {year} {2021})}\BibitemShut {NoStop}%
\bibitem [{\citenamefont {Mandle}(2022{\natexlab{a}})}]{Mandle2022polar}%
  \BibitemOpen
  \bibfield  {author} {\bibinfo {author} {\bibfnamefont {R.~J.}\ \bibnamefont {Mandle}},\ }\bibfield  {title} {\bibinfo {title} {A new order of liquids: polar order in nematic liquid crystals},\ }\href@noop {} {\bibfield  {journal} {\bibinfo  {journal} {Soft Matter}\ }\textbf {\bibinfo {volume} {18}},\ \bibinfo {pages} {5014} (\bibinfo {year} {2022}{\natexlab{a}})}\BibitemShut {NoStop}%
\bibitem [{\citenamefont {Fernández-Rico}\ \emph {et~al.}(2020{\natexlab{a}})\citenamefont {Fernández-Rico}, \citenamefont {Chiappini}, \citenamefont {Yanagishima}, \citenamefont {de~Sousa}, \citenamefont {Aarts}, \citenamefont {Dijkstra},\ and\ \citenamefont {Dullens}}]{Fernandez-Rico2020}%
  \BibitemOpen
  \bibfield  {author} {\bibinfo {author} {\bibfnamefont {C.}~\bibnamefont {Fernández-Rico}}, \bibinfo {author} {\bibfnamefont {M.}~\bibnamefont {Chiappini}}, \bibinfo {author} {\bibfnamefont {T.}~\bibnamefont {Yanagishima}}, \bibinfo {author} {\bibfnamefont {H.}~\bibnamefont {de~Sousa}}, \bibinfo {author} {\bibfnamefont {D.~G. A.~L.}\ \bibnamefont {Aarts}}, \bibinfo {author} {\bibfnamefont {M.}~\bibnamefont {Dijkstra}},\ and\ \bibinfo {author} {\bibfnamefont {R.~P.~A.}\ \bibnamefont {Dullens}},\ }\bibfield  {title} {\bibinfo {title} {Shaping colloidal bananas to reveal biaxial, splay-bend nematic, and smectic phases},\ }\href {https://doi.org/10.1126/science.abb4536} {\bibfield  {journal} {\bibinfo  {journal} {Science}\ }\textbf {\bibinfo {volume} {369}},\ \bibinfo {pages} {950} (\bibinfo {year} {2020}{\natexlab{a}})}\BibitemShut {NoStop}%
\bibitem [{\citenamefont {Kotni}\ \emph {et~al.}(2022)\citenamefont {Kotni}, \citenamefont {Grau-Carbonell}, \citenamefont {Chiappini}, \citenamefont {Dijkstra},\ and\ \citenamefont {van Blaaderen}}]{Kotni2022}%
  \BibitemOpen
  \bibfield  {author} {\bibinfo {author} {\bibfnamefont {R.}~\bibnamefont {Kotni}}, \bibinfo {author} {\bibfnamefont {A.}~\bibnamefont {Grau-Carbonell}}, \bibinfo {author} {\bibfnamefont {M.}~\bibnamefont {Chiappini}}, \bibinfo {author} {\bibfnamefont {M.}~\bibnamefont {Dijkstra}},\ and\ \bibinfo {author} {\bibfnamefont {A.}~\bibnamefont {van Blaaderen}},\ }\bibfield  {title} {\bibinfo {title} {Splay-bend nematic phases of bent colloidal silica rods induced by polydispersity},\ }\href {https://doi.org/10.1038/s41467-022-34658-y} {\bibfield  {journal} {\bibinfo  {journal} {Nature Communications}\ }\textbf {\bibinfo {volume} {13}},\ \bibinfo {pages} {7264} (\bibinfo {year} {2022})}\BibitemShut {NoStop}%
\bibitem [{\citenamefont {Mertelj}\ \emph {et~al.}(2018{\natexlab{b}})\citenamefont {Mertelj}, \citenamefont {Cmok}, \citenamefont {Sebastián}, \citenamefont {Mandle}, \citenamefont {Parker}, \citenamefont {Whitwood}, \citenamefont {Goodby},\ and\ \citenamefont {Čopič}}]{Mertelj2018}%
  \BibitemOpen
  \bibfield  {author} {\bibinfo {author} {\bibfnamefont {A.}~\bibnamefont {Mertelj}}, \bibinfo {author} {\bibfnamefont {L.}~\bibnamefont {Cmok}}, \bibinfo {author} {\bibfnamefont {N.}~\bibnamefont {Sebastián}}, \bibinfo {author} {\bibfnamefont {R.~J.}\ \bibnamefont {Mandle}}, \bibinfo {author} {\bibfnamefont {R.~R.}\ \bibnamefont {Parker}}, \bibinfo {author} {\bibfnamefont {A.~C.}\ \bibnamefont {Whitwood}}, \bibinfo {author} {\bibfnamefont {J.~W.}\ \bibnamefont {Goodby}},\ and\ \bibinfo {author} {\bibfnamefont {M.}~\bibnamefont {Čopič}},\ }\bibfield  {title} {\bibinfo {title} {Splay nematic phase},\ }\href {https://doi.org/10.1103/PhysRevX.8.041025} {\bibfield  {journal} {\bibinfo  {journal} {Physical Review X}\ }\textbf {\bibinfo {volume} {8}},\ \bibinfo {pages} {041025} (\bibinfo {year} {2018}{\natexlab{b}})}\BibitemShut {NoStop}%
\bibitem [{\citenamefont {Chen}\ \emph {et~al.}(2020{\natexlab{b}})\citenamefont {Chen}, \citenamefont {Korblova}, \citenamefont {Dong}, \citenamefont {Wei}, \citenamefont {Shao}, \citenamefont {Radzihovsky}, \citenamefont {Glaser}, \citenamefont {Maclennan}, \citenamefont {Bedrov}, \citenamefont {Walba},\ and\ \citenamefont {Clark}}]{Chen2020}%
  \BibitemOpen
  \bibfield  {author} {\bibinfo {author} {\bibfnamefont {X.}~\bibnamefont {Chen}}, \bibinfo {author} {\bibfnamefont {E.}~\bibnamefont {Korblova}}, \bibinfo {author} {\bibfnamefont {D.}~\bibnamefont {Dong}}, \bibinfo {author} {\bibfnamefont {X.}~\bibnamefont {Wei}}, \bibinfo {author} {\bibfnamefont {R.}~\bibnamefont {Shao}}, \bibinfo {author} {\bibfnamefont {L.}~\bibnamefont {Radzihovsky}}, \bibinfo {author} {\bibfnamefont {M.~A.}\ \bibnamefont {Glaser}}, \bibinfo {author} {\bibfnamefont {J.~E.}\ \bibnamefont {Maclennan}}, \bibinfo {author} {\bibfnamefont {D.}~\bibnamefont {Bedrov}}, \bibinfo {author} {\bibfnamefont {D.~M.}\ \bibnamefont {Walba}},\ and\ \bibinfo {author} {\bibfnamefont {N.~A.}\ \bibnamefont {Clark}},\ }\bibfield  {title} {\bibinfo {title} {First-principles experimental demonstration of ferroelectricity in a thermotropic nematic liquid crystal: Polar domains and striking electro-optics},\ }\href {https://doi.org/10.1073/pnas.2002290117} {\bibfield  {journal} {\bibinfo  {journal} {Proceedings of
  the National Academy of Sciences}\ }\textbf {\bibinfo {volume} {117}},\ \bibinfo {pages} {14021} (\bibinfo {year} {2020}{\natexlab{b}})}\BibitemShut {NoStop}%
\bibitem [{\citenamefont {J\'akli}\ \emph {et~al.}(2018)\citenamefont {J\'akli}, \citenamefont {Lavrentovich},\ and\ \citenamefont {Selinger}}]{RevModPhys.90.045004}%
  \BibitemOpen
  \bibfield  {author} {\bibinfo {author} {\bibfnamefont {A.}~\bibnamefont {J\'akli}}, \bibinfo {author} {\bibfnamefont {O.~D.}\ \bibnamefont {Lavrentovich}},\ and\ \bibinfo {author} {\bibfnamefont {J.~V.}\ \bibnamefont {Selinger}},\ }\bibfield  {title} {\bibinfo {title} {Physics of liquid crystals of bent-shaped molecules},\ }\href {https://doi.org/10.1103/RevModPhys.90.045004} {\bibfield  {journal} {\bibinfo  {journal} {Rev. Mod. Phys.}\ }\textbf {\bibinfo {volume} {90}},\ \bibinfo {pages} {045004} (\bibinfo {year} {2018})}\BibitemShut {NoStop}%
\bibitem [{\citenamefont {Mandle}(2022{\natexlab{b}})}]{Mandle2022}%
  \BibitemOpen
  \bibfield  {author} {\bibinfo {author} {\bibfnamefont {R.~J.}\ \bibnamefont {Mandle}},\ }\bibfield  {title} {\bibinfo {title} {A ten-year perspective on twist-bend nematic materials},\ }\href {https://doi.org/10.3390/molecules27092689} {\bibfield  {journal} {\bibinfo  {journal} {Molecules}\ }\textbf {\bibinfo {volume} {27}},\ \bibinfo {pages} {2689} (\bibinfo {year} {2022}{\natexlab{b}})}\BibitemShut {NoStop}%
\bibitem [{\citenamefont {B.}(1976)}]{MeyerFlexopolarization}%
  \BibitemOpen
  \bibfield  {author} {\bibinfo {author} {\bibfnamefont {M.~R.}\ \bibnamefont {B.}},\ }\href@noop {} {\emph {\bibinfo {title} {Molecular Fluids}}},\ edited by\ \bibinfo {editor} {\bibfnamefont {R.}~\bibnamefont {Balian}}\ and\ \bibinfo {editor} {\bibfnamefont {G.}~\bibnamefont {Weill}},\ Proceedings of the Les Houches Summer School on Theoretical Physics, 1973, session No. XXV\ (\bibinfo  {publisher} {Gordon and Breach Science Publishers},\ \bibinfo {year} {1976})\ p.\ \bibinfo {pages} {271}\BibitemShut {NoStop}%
\bibitem [{\citenamefont {Longa}\ and\ \citenamefont {Trebin}(1990)}]{Longa1990}%
  \BibitemOpen
  \bibfield  {author} {\bibinfo {author} {\bibfnamefont {L.}~\bibnamefont {Longa}}\ and\ \bibinfo {author} {\bibfnamefont {H.-R.}\ \bibnamefont {Trebin}},\ }\bibfield  {title} {\bibinfo {title} {Spontaneous polarization in chiral biaxial liquid crystals},\ }\href {https://doi.org/10.1103/PhysRevA.42.3453} {\bibfield  {journal} {\bibinfo  {journal} {Physical Review A}\ }\textbf {\bibinfo {volume} {42}},\ \bibinfo {pages} {3453} (\bibinfo {year} {1990})}\BibitemShut {NoStop}%
\bibitem [{\citenamefont {Longa}\ and\ \citenamefont {Tomczyk}(2020)}]{Longa2020}%
  \BibitemOpen
  \bibfield  {author} {\bibinfo {author} {\bibfnamefont {L.}~\bibnamefont {Longa}}\ and\ \bibinfo {author} {\bibfnamefont {W.}~\bibnamefont {Tomczyk}},\ }\bibfield  {title} {\bibinfo {title} {Twist–{B}end {N}ematic {P}hase from the {L}andau–de {G}ennes {P}erspective},\ }\href {https://doi.org/10.1021/acs.jpcc.0c05711} {\bibfield  {journal} {\bibinfo  {journal} {J. Phys. Chem. C}\ }\textbf {\bibinfo {volume} {124}},\ \bibinfo {pages} {22761} (\bibinfo {year} {2020})}\BibitemShut {NoStop}%
\bibitem [{\citenamefont {Dozov}(2001)}]{Dozov_2001}%
  \BibitemOpen
  \bibfield  {author} {\bibinfo {author} {\bibfnamefont {I.}~\bibnamefont {Dozov}},\ }\bibfield  {title} {\bibinfo {title} {On the spontaneous symmetry breaking in the mesophases of achiral banana-shaped molecules},\ }\href {https://doi.org/10.1209/epl/i2001-00513-x} {\bibfield  {journal} {\bibinfo  {journal} {Europhysics Letters ({EPL})}\ }\textbf {\bibinfo {volume} {56}},\ \bibinfo {pages} {247} (\bibinfo {year} {2001})}\BibitemShut {NoStop}%
\bibitem [{\citenamefont {Shamid}\ \emph {et~al.}(2013)\citenamefont {Shamid}, \citenamefont {Dhakal},\ and\ \citenamefont {Selinger}}]{Shamid2013statistical}%
  \BibitemOpen
  \bibfield  {author} {\bibinfo {author} {\bibfnamefont {S.~M.}\ \bibnamefont {Shamid}}, \bibinfo {author} {\bibfnamefont {S.}~\bibnamefont {Dhakal}},\ and\ \bibinfo {author} {\bibfnamefont {J.~V.}\ \bibnamefont {Selinger}},\ }\bibfield  {title} {\bibinfo {title} {Statistical mechanics of bend flexoelectricity and the twist-bend phase in bent-core liquid crystals},\ }\href {https://doi.org/10.1103/PhysRevE.87.052503} {\bibfield  {journal} {\bibinfo  {journal} {Phys. Rev. E}\ }\textbf {\bibinfo {volume} {87}},\ \bibinfo {pages} {052503} (\bibinfo {year} {2013})}\BibitemShut {NoStop}%
\bibitem [{\citenamefont {Lorman}\ and\ \citenamefont {Mettout}(2004)}]{LormanMettout2004}%
  \BibitemOpen
  \bibfield  {author} {\bibinfo {author} {\bibfnamefont {V.~L.}\ \bibnamefont {Lorman}}\ and\ \bibinfo {author} {\bibfnamefont {B.}~\bibnamefont {Mettout}},\ }\bibfield  {title} {\bibinfo {title} {Theory of chiral periodic mesophases formed from an achiral liquid of bent-core molecules},\ }\href {https://doi.org/10.1103/PhysRevE.69.061710} {\bibfield  {journal} {\bibinfo  {journal} {Phys. Rev. E}\ }\textbf {\bibinfo {volume} {69}},\ \bibinfo {pages} {061710} (\bibinfo {year} {2004})}\BibitemShut {NoStop}%
\bibitem [{\citenamefont {Shamid}\ \emph {et~al.}(2014)\citenamefont {Shamid}, \citenamefont {Allender},\ and\ \citenamefont {Selinger}}]{Shamid2014}%
  \BibitemOpen
  \bibfield  {author} {\bibinfo {author} {\bibfnamefont {S.~M.}\ \bibnamefont {Shamid}}, \bibinfo {author} {\bibfnamefont {D.~W.}\ \bibnamefont {Allender}},\ and\ \bibinfo {author} {\bibfnamefont {J.~V.}\ \bibnamefont {Selinger}},\ }\bibfield  {title} {\bibinfo {title} {Predicting a polar analog of chiral blue phases in liquid crystals},\ }\href {https://doi.org/10.1103/PhysRevLett.113.237801} {\bibfield  {journal} {\bibinfo  {journal} {Phys. Rev. Lett.}\ }\textbf {\bibinfo {volume} {113}},\ \bibinfo {pages} {237801} (\bibinfo {year} {2014})}\BibitemShut {NoStop}%
\bibitem [{\citenamefont {Longa}\ and\ \citenamefont {Paj\c{a}k}(2016)}]{Longa2016}%
  \BibitemOpen
  \bibfield  {author} {\bibinfo {author} {\bibfnamefont {L.}~\bibnamefont {Longa}}\ and\ \bibinfo {author} {\bibfnamefont {G.}~\bibnamefont {Paj\c{a}k}},\ }\bibfield  {title} {\bibinfo {title} {Modulated nematic structures induced by chirality and steric polarization},\ }\href {https://doi.org/10.1103/PhysRevE.93.040701} {\bibfield  {journal} {\bibinfo  {journal} {Phys. Rev. E}\ }\textbf {\bibinfo {volume} {93}},\ \bibinfo {pages} {040701} (\bibinfo {year} {2016})}\BibitemShut {NoStop}%
\bibitem [{\citenamefont {Pająk}\ \emph {et~al.}(2018)\citenamefont {Pająk}, \citenamefont {Longa},\ and\ \citenamefont {Chrzanowska}}]{Pajak2018}%
  \BibitemOpen
  \bibfield  {author} {\bibinfo {author} {\bibfnamefont {G.}~\bibnamefont {Pająk}}, \bibinfo {author} {\bibfnamefont {L.}~\bibnamefont {Longa}},\ and\ \bibinfo {author} {\bibfnamefont {A.}~\bibnamefont {Chrzanowska}},\ }\bibfield  {title} {\bibinfo {title} {Nematic twist-bend phase in an external field},\ }\href {https://doi.org/10.1073/pnas.1721786115} {\bibfield  {journal} {\bibinfo  {journal} {Proceedings of the National Academy of Sciences}\ }\textbf {\bibinfo {volume} {115}},\ \bibinfo {pages} {E10303} (\bibinfo {year} {2018})}\BibitemShut {NoStop}%
\bibitem [{\citenamefont {Fernández-Rico}\ \emph {et~al.}(2020{\natexlab{b}})\citenamefont {Fernández-Rico}, \citenamefont {Chiappini}, \citenamefont {Yanagishima}, \citenamefont {de~Sousa}, \citenamefont {Aarts}, \citenamefont {Dijkstra},\ and\ \citenamefont {Dullens}}]{Rico2020}%
  \BibitemOpen
  \bibfield  {author} {\bibinfo {author} {\bibfnamefont {C.}~\bibnamefont {Fernández-Rico}}, \bibinfo {author} {\bibfnamefont {M.}~\bibnamefont {Chiappini}}, \bibinfo {author} {\bibfnamefont {T.}~\bibnamefont {Yanagishima}}, \bibinfo {author} {\bibfnamefont {H.}~\bibnamefont {de~Sousa}}, \bibinfo {author} {\bibfnamefont {D.~G. A.~L.}\ \bibnamefont {Aarts}}, \bibinfo {author} {\bibfnamefont {M.}~\bibnamefont {Dijkstra}},\ and\ \bibinfo {author} {\bibfnamefont {R.~P.~A.}\ \bibnamefont {Dullens}},\ }\bibfield  {title} {\bibinfo {title} {Shaping colloidal bananas to reveal biaxial, splay-bend nematic, and smectic phases},\ }\href {https://doi.org/10.1126/science.abb4536} {\bibfield  {journal} {\bibinfo  {journal} {Science}\ }\textbf {\bibinfo {volume} {369}},\ \bibinfo {pages} {950} (\bibinfo {year} {2020}{\natexlab{b}})}\BibitemShut {NoStop}%
\bibitem [{\citenamefont {Chiappini}\ and\ \citenamefont {Dijkstra}(2021)}]{Chiappini2021}%
  \BibitemOpen
  \bibfield  {author} {\bibinfo {author} {\bibfnamefont {M.}~\bibnamefont {Chiappini}}\ and\ \bibinfo {author} {\bibfnamefont {M.}~\bibnamefont {Dijkstra}},\ }\bibfield  {title} {\bibinfo {title} {A generalized density-modulated twist-splay-bend phase of banana-shaped particles},\ }\href {https://doi.org/10.1038/s41467-021-22413-8} {\bibfield  {journal} {\bibinfo  {journal} {Nat. Commun.}\ }\textbf {\bibinfo {volume} {12}},\ \bibinfo {pages} {2157} (\bibinfo {year} {2021})}\BibitemShut {NoStop}%
\bibitem [{\citenamefont {Kubala}\ \emph {et~al.}(2022)\citenamefont {Kubala}, \citenamefont {Tomczyk},\ and\ \citenamefont {Cie{\'{s}}la}}]{Kubala2022}%
  \BibitemOpen
  \bibfield  {author} {\bibinfo {author} {\bibfnamefont {P.}~\bibnamefont {Kubala}}, \bibinfo {author} {\bibfnamefont {W.}~\bibnamefont {Tomczyk}},\ and\ \bibinfo {author} {\bibfnamefont {M.}~\bibnamefont {Cie{\'{s}}la}},\ }\bibfield  {title} {\bibinfo {title} {{In silico study of liquid crystalline phases formed by bent-shaped molecules with excluded volume type interactions}},\ }\href {https://doi.org/10.1016/j.molliq.2022.120156} {\bibfield  {journal} {\bibinfo  {journal} {J. Mol. Liq.}\ }\textbf {\bibinfo {volume} {367}},\ \bibinfo {pages} {120156} (\bibinfo {year} {2022})}\BibitemShut {NoStop}%
\bibitem [{\citenamefont {Dunmur}(2022)}]{Dunmur2022}%
  \BibitemOpen
  \bibfield  {author} {\bibinfo {author} {\bibfnamefont {D.}~\bibnamefont {Dunmur}},\ }\bibfield  {title} {\bibinfo {title} {Anatomy of a discovery: The twist–bend nematic phase},\ }\bibfield  {journal} {\bibinfo  {journal} {Crystals}\ }\textbf {\bibinfo {volume} {12}},\ \href {https://doi.org/10.3390/cryst12030309} {10.3390/cryst12030309} (\bibinfo {year} {2022})\BibitemShut {NoStop}%
\bibitem [{\citenamefont {Meyer}\ \emph {et~al.}(2020)\citenamefont {Meyer}, \citenamefont {Blanc}, \citenamefont {Luckhurst}, \citenamefont {Davidson},\ and\ \citenamefont {Dozov}}]{Meyer2020}%
  \BibitemOpen
  \bibfield  {author} {\bibinfo {author} {\bibfnamefont {C.}~\bibnamefont {Meyer}}, \bibinfo {author} {\bibfnamefont {C.}~\bibnamefont {Blanc}}, \bibinfo {author} {\bibfnamefont {G.~R.}\ \bibnamefont {Luckhurst}}, \bibinfo {author} {\bibfnamefont {P.}~\bibnamefont {Davidson}},\ and\ \bibinfo {author} {\bibfnamefont {I.}~\bibnamefont {Dozov}},\ }\bibfield  {title} {\bibinfo {title} {Biaxiality-driven twist-bend to splay-bend nematic phase transition induced by an electric field},\ }\href {https://doi.org/10.1126/sciadv.abb8212} {\bibfield  {journal} {\bibinfo  {journal} {Science Advances}\ }\textbf {\bibinfo {volume} {6}},\ \bibinfo {pages} {eabb8212} (\bibinfo {year} {2020})}\BibitemShut {NoStop}%
\bibitem [{\citenamefont {Merkel}\ \emph {et~al.}(2018)\citenamefont {Merkel}, \citenamefont {Kocot}, \citenamefont {Vij},\ and\ \citenamefont {Shanker}}]{Merkel2018}%
  \BibitemOpen
  \bibfield  {author} {\bibinfo {author} {\bibfnamefont {K.}~\bibnamefont {Merkel}}, \bibinfo {author} {\bibfnamefont {A.}~\bibnamefont {Kocot}}, \bibinfo {author} {\bibfnamefont {J.~K.}\ \bibnamefont {Vij}},\ and\ \bibinfo {author} {\bibfnamefont {G.}~\bibnamefont {Shanker}},\ }\bibfield  {title} {\bibinfo {title} {Distortions in structures of the twist bend nematic phase of a bent-core liquid crystal by the electric field},\ }\href {https://doi.org/10.1103/PhysRevE.98.022704} {\bibfield  {journal} {\bibinfo  {journal} {Physical Review E}\ }\textbf {\bibinfo {volume} {98}},\ \bibinfo {pages} {022704} (\bibinfo {year} {2018})}\BibitemShut {NoStop}%
\bibitem [{\citenamefont {Panov}\ \emph {et~al.}(2021)\citenamefont {Panov}, \citenamefont {Vij},\ and\ \citenamefont {Mehl}}]{Panov2021}%
  \BibitemOpen
  \bibfield  {author} {\bibinfo {author} {\bibfnamefont {V.~P.}\ \bibnamefont {Panov}}, \bibinfo {author} {\bibfnamefont {J.~K.}\ \bibnamefont {Vij}},\ and\ \bibinfo {author} {\bibfnamefont {G.~H.}\ \bibnamefont {Mehl}},\ }\bibfield  {title} {\bibinfo {title} {The beauty of twist-bend nematic phase: Fast switching domains, first order fréedericksz transition and a hierarchy of structures},\ }\href {https://doi.org/10.3390/cryst11060621} {\bibfield  {journal} {\bibinfo  {journal} {Crystals}\ }\textbf {\bibinfo {volume} {11}},\ \bibinfo {pages} {621} (\bibinfo {year} {2021})}\BibitemShut {NoStop}%
\bibitem [{\citenamefont {Weeks}\ \emph {et~al.}(1971)\citenamefont {Weeks}, \citenamefont {Chandler},\ and\ \citenamefont {Andersen}}]{WCA1971}%
  \BibitemOpen
  \bibfield  {author} {\bibinfo {author} {\bibfnamefont {J.~D.}\ \bibnamefont {Weeks}}, \bibinfo {author} {\bibfnamefont {D.}~\bibnamefont {Chandler}},\ and\ \bibinfo {author} {\bibfnamefont {H.~C.}\ \bibnamefont {Andersen}},\ }\bibfield  {title} {\bibinfo {title} {Role of repulsive forces in determining the equilibrium structure of simple liquids},\ }\href {https://doi.org/10.1063/1.1674820} {\bibfield  {journal} {\bibinfo  {journal} {J. Chem. Phys.}\ }\textbf {\bibinfo {volume} {54}},\ \bibinfo {pages} {5237} (\bibinfo {year} {1971})}\BibitemShut {NoStop}%
\bibitem [{\citenamefont {Chandler}\ \emph {et~al.}(1983)\citenamefont {Chandler}, \citenamefont {Weeks},\ and\ \citenamefont {Andersen}}]{WCA1983}%
  \BibitemOpen
  \bibfield  {author} {\bibinfo {author} {\bibfnamefont {D.}~\bibnamefont {Chandler}}, \bibinfo {author} {\bibfnamefont {J.~D.}\ \bibnamefont {Weeks}},\ and\ \bibinfo {author} {\bibfnamefont {H.~C.}\ \bibnamefont {Andersen}},\ }\bibfield  {title} {\bibinfo {title} {Van der waals picture of liquids, solids, and phase transformations},\ }\href {https://doi.org/10.1126/science.220.4599.787} {\bibfield  {journal} {\bibinfo  {journal} {Science}\ }\textbf {\bibinfo {volume} {220}},\ \bibinfo {pages} {787} (\bibinfo {year} {1983})}\BibitemShut {NoStop}%
\bibitem [{\citenamefont {Heyes}\ and\ \citenamefont {Okumura}(2006)}]{Heyes2006}%
  \BibitemOpen
  \bibfield  {author} {\bibinfo {author} {\bibfnamefont {D.~M.}\ \bibnamefont {Heyes}}\ and\ \bibinfo {author} {\bibfnamefont {H.}~\bibnamefont {Okumura}},\ }\bibfield  {title} {\bibinfo {title} {Equation of state and structural properties of the {W}eeks-{C}handler-{A}ndersen fluid},\ }\href@noop {} {\bibfield  {journal} {\bibinfo  {journal} {J. Chem. Phys.}\ }\textbf {\bibinfo {volume} {124}},\ \bibinfo {pages} {164507} (\bibinfo {year} {2006})}\BibitemShut {NoStop}%
\bibitem [{\citenamefont {Greco}\ and\ \citenamefont {Ferrarini}(2015)}]{GrecoFerrariniPRL}%
  \BibitemOpen
  \bibfield  {author} {\bibinfo {author} {\bibfnamefont {C.}~\bibnamefont {Greco}}\ and\ \bibinfo {author} {\bibfnamefont {A.}~\bibnamefont {Ferrarini}},\ }\bibfield  {title} {\bibinfo {title} {Entropy-driven chiral order in a system of achiral bent particles},\ }\href {https://doi.org/10.1103/PhysRevLett.115.147801} {\bibfield  {journal} {\bibinfo  {journal} {Phys. Rev. Lett.}\ }\textbf {\bibinfo {volume} {115}},\ \bibinfo {pages} {147801} (\bibinfo {year} {2015})}\BibitemShut {NoStop}%
\bibitem [{\citenamefont {Thompson}\ \emph {et~al.}(2022)\citenamefont {Thompson}, \citenamefont {Aktulga}, \citenamefont {Berger}, \citenamefont {Bolintineanu}, \citenamefont {Brown}, \citenamefont {Crozier}, \citenamefont {in~'t Veld}, \citenamefont {Kohlmeyer}, \citenamefont {Moore}, \citenamefont {Nguyen}, \citenamefont {Shan}, \citenamefont {Stevens}, \citenamefont {Tranchida}, \citenamefont {Trott},\ and\ \citenamefont {Plimpton}}]{LAMMPS}%
  \BibitemOpen
  \bibfield  {author} {\bibinfo {author} {\bibfnamefont {A.~P.}\ \bibnamefont {Thompson}}, \bibinfo {author} {\bibfnamefont {H.~M.}\ \bibnamefont {Aktulga}}, \bibinfo {author} {\bibfnamefont {R.}~\bibnamefont {Berger}}, \bibinfo {author} {\bibfnamefont {D.~S.}\ \bibnamefont {Bolintineanu}}, \bibinfo {author} {\bibfnamefont {W.~M.}\ \bibnamefont {Brown}}, \bibinfo {author} {\bibfnamefont {P.~S.}\ \bibnamefont {Crozier}}, \bibinfo {author} {\bibfnamefont {P.~J.}\ \bibnamefont {in~'t Veld}}, \bibinfo {author} {\bibfnamefont {A.}~\bibnamefont {Kohlmeyer}}, \bibinfo {author} {\bibfnamefont {S.~G.}\ \bibnamefont {Moore}}, \bibinfo {author} {\bibfnamefont {T.~D.}\ \bibnamefont {Nguyen}}, \bibinfo {author} {\bibfnamefont {R.}~\bibnamefont {Shan}}, \bibinfo {author} {\bibfnamefont {M.~J.}\ \bibnamefont {Stevens}}, \bibinfo {author} {\bibfnamefont {J.}~\bibnamefont {Tranchida}}, \bibinfo {author} {\bibfnamefont {C.}~\bibnamefont {Trott}},\ and\ \bibinfo {author} {\bibfnamefont {S.~J.}\ \bibnamefont {Plimpton}},\
  }\bibfield  {title} {\bibinfo {title} {{LAMMPS} - a flexible simulation tool for particle-based materials modeling at the atomic, meso, and continuum scales},\ }\href {https://doi.org/10.1016/j.cpc.2021.108171} {\bibfield  {journal} {\bibinfo  {journal} {Comp. Phys. Comm.}\ }\textbf {\bibinfo {volume} {271}},\ \bibinfo {pages} {108171} (\bibinfo {year} {2022})}\BibitemShut {NoStop}%
\bibitem [{\citenamefont {Xia}\ \emph {et~al.}(2019)\citenamefont {Xia}, \citenamefont {DeBenedictis}, \citenamefont {Kim}, \citenamefont {Chen}, \citenamefont {Kim}, \citenamefont {Cleaver}, \citenamefont {Atherton},\ and\ \citenamefont {Yang}}]{XiaNatCommun2019}%
  \BibitemOpen
  \bibfield  {author} {\bibinfo {author} {\bibfnamefont {Y.}~\bibnamefont {Xia}}, \bibinfo {author} {\bibfnamefont {A.~A.}\ \bibnamefont {DeBenedictis}}, \bibinfo {author} {\bibfnamefont {D.~S.}\ \bibnamefont {Kim}}, \bibinfo {author} {\bibfnamefont {S.}~\bibnamefont {Chen}}, \bibinfo {author} {\bibfnamefont {S.-U.}\ \bibnamefont {Kim}}, \bibinfo {author} {\bibfnamefont {D.~J.}\ \bibnamefont {Cleaver}}, \bibinfo {author} {\bibfnamefont {T.~J.}\ \bibnamefont {Atherton}},\ and\ \bibinfo {author} {\bibfnamefont {S.}~\bibnamefont {Yang}},\ }\bibfield  {title} {\bibinfo {title} {Programming emergent symmetries with saddle-splay elasticity},\ }\href {https://doi.org/10.1038/s41467-019-13012-9} {\bibfield  {journal} {\bibinfo  {journal} {Nature Communications}\ }\textbf {\bibinfo {volume} {10}},\ \bibinfo {pages} {5104} (\bibinfo {year} {2019})}\BibitemShut {NoStop}%
\bibitem [{\citenamefont {Selinger}(2018)}]{SelingerLCR2018}%
  \BibitemOpen
  \bibfield  {author} {\bibinfo {author} {\bibfnamefont {J.~V.}\ \bibnamefont {Selinger}},\ }\bibfield  {title} {\bibinfo {title} {Interpretation of saddle-splay and the oseen--frank free energy in liquid crystals},\ }\href {https://doi.org/10.1080/21680396.2019.1581103} {\bibfield  {journal} {\bibinfo  {journal} {Liquid Crystals Reviews}\ }\textbf {\bibinfo {volume} {6}},\ \bibinfo {pages} {129} (\bibinfo {year} {2018})},\ \bibinfo {note} {published online: 1 March 2019}\BibitemShut {NoStop}%
\bibitem [{\citenamefont {Sobel}\ and\ \citenamefont {Feldman}(1968)}]{Sobel1968}%
  \BibitemOpen
  \bibfield  {author} {\bibinfo {author} {\bibfnamefont {I.}~\bibnamefont {Sobel}}\ and\ \bibinfo {author} {\bibfnamefont {G.}~\bibnamefont {Feldman}},\ }\href {https://www.researchgate.net/publication/281104656_An_Isotropic_3x3_Image_Gradient_Operator} {\bibinfo {title} {A 3{\texttimes}3 isotropic gradient operator for image processing}},\ \bibinfo {howpublished} {Presented at the Stanford Artificial Intelligence Project (SAIL)} (\bibinfo {year} {1968}),\ \bibinfo {note} {historical write-up: ``An Isotropic 3{\texttimes}3 Image Gradient Operator'' (2014) by I. Sobel}\BibitemShut {NoStop}%
\bibitem [{\citenamefont {Gonzalez}\ and\ \citenamefont {Woods}(2006)}]{GonzalezWoods2006}%
  \BibitemOpen
  \bibfield  {author} {\bibinfo {author} {\bibfnamefont {R.~C.}\ \bibnamefont {Gonzalez}}\ and\ \bibinfo {author} {\bibfnamefont {R.~E.}\ \bibnamefont {Woods}},\ }\href {https://dl.acm.org/doi/10.5555/1076432} {\emph {\bibinfo {title} {Digital Image Processing}}},\ \bibinfo {edition} {3rd}\ ed.\ (\bibinfo  {publisher} {Prentice Hall},\ \bibinfo {address} {Upper Saddle River, NJ},\ \bibinfo {year} {2006})\BibitemShut {NoStop}%
\bibitem [{\citenamefont {Longa}\ \emph {et~al.}(2023)\citenamefont {Longa}, \citenamefont {Cie{\'s}la}, \citenamefont {Karbowniczek},\ and\ \citenamefont {Chrzanowska}}]{Longa2023}%
  \BibitemOpen
  \bibfield  {author} {\bibinfo {author} {\bibfnamefont {L.}~\bibnamefont {Longa}}, \bibinfo {author} {\bibfnamefont {M.}~\bibnamefont {Cie{\'s}la}}, \bibinfo {author} {\bibfnamefont {P.}~\bibnamefont {Karbowniczek}},\ and\ \bibinfo {author} {\bibfnamefont {A.}~\bibnamefont {Chrzanowska}},\ }\bibfield  {title} {\bibinfo {title} {Conformational degrees of freedom and stability of splay-bend ordering in the limit of a very strong planar anchoring},\ }\href {https://doi.org/10.1103/PhysRevE.107.034707} {\bibfield  {journal} {\bibinfo  {journal} {Physical Review E}\ }\textbf {\bibinfo {volume} {107}},\ \bibinfo {pages} {034707} (\bibinfo {year} {2023})}\BibitemShut {NoStop}%
\bibitem [{\citenamefont {Karbowniczek}\ \emph {et~al.}(2017)\citenamefont {Karbowniczek}, \citenamefont {Cieśla}, \citenamefont {Longa},\ and\ \citenamefont {Chrzanowska}}]{Karbowniczek2017}%
  \BibitemOpen
  \bibfield  {author} {\bibinfo {author} {\bibfnamefont {P.}~\bibnamefont {Karbowniczek}}, \bibinfo {author} {\bibfnamefont {M.}~\bibnamefont {Cieśla}}, \bibinfo {author} {\bibfnamefont {L.}~\bibnamefont {Longa}},\ and\ \bibinfo {author} {\bibfnamefont {A.}~\bibnamefont {Chrzanowska}},\ }\bibfield  {title} {\bibinfo {title} {Structure formation in monolayers composed of hard bent-core molecules},\ }\href {https://doi.org/10.1080/02678292.2016.1259510} {\bibfield  {journal} {\bibinfo  {journal} {Liq. Cryst.}\ }\textbf {\bibinfo {volume} {44}},\ \bibinfo {pages} {254} (\bibinfo {year} {2017})}\BibitemShut {NoStop}%
\end{thebibliography}%

\end{document}